\documentclass[twocolumn]{aastex631}
\usepackage{amsmath}
\usepackage{amsthm}
\usepackage{amssymb}
\usepackage{graphicx}
\usepackage{physics}
\usepackage{color}
\usepackage{hyperref}
\usepackage{bm}
\usepackage{footmisc}
\newcommand*\boldell{\pmb{\ell}}

\shorttitle{kSZ$^2$--galaxy Cross-Correlations during Reionization}
\shortauthors{La Plante, Sipple, \& Lidz}

\submitjournal{ApJ}

\begin{document}

\defcitealias{waters_etal2016}{W16}
\defcitealias{ferraro_etal2016}{F16}

\title{Prospects for kSZ$^2$--Galaxy Cross-Correlations during Reionization}

\author[0000-0002-4693-0102]{Paul La Plante}
\affiliation{Astronomy Department, University of California, Berkeley, CA 94720, USA}
\affiliation{Berkeley Center for Cosmological Physics, University of California, Berkeley, CA 94720, USA}

\author{Jackson Sipple}
\affiliation{Center for Particle Cosmology, Department of Physics and Astronomy, University of Pennsylvania, Philadelphia, PA 19104, USA}

\author[0000-0002-3950-9598]{Adam Lidz}
\affiliation{Center for Particle Cosmology, Department of Physics and Astronomy, University of Pennsylvania, Philadelphia, PA 19104, USA}

\correspondingauthor{Paul La Plante}
\email{plaplant@berkeley.edu}

\begin{abstract}
  We explore a new approach for extracting reionization-era contributions to the
  kinetic Sunyaev--Zel'dovich (kSZ) effect. Our method utilizes the cross-power
  spectrum between filtered and squared maps of the cosmic microwave background
  (CMB) and photometric galaxy surveys during the Epoch of Reionization
  (EoR). This kSZ$^2$--galaxy cross-power spectrum statistic has been
  successfully detected at lower redshifts ($z \lesssim 1.5$). Here we extend
  this method to $z \gtrsim 6$ as a potential means to extract signatures of
  patchy reionization. We model the expected signal across multiple photometric
  redshift bins using seminumeric simulations of the reionization process. In
  principle, the cross-correlation statistic robustly extracts reionization-era
  contributions to the kSZ signal, while its redshift evolution yields valuable
  information regarding the timing of reionization. Specifically, the model
  cross-correlation signal near $\ell \sim 1000$ peaks during the early stages
  of the EoR, when about 20\% of the volume of the universe is
  ionized. Detectable $\ell$ modes mainly reflect squeezed-triangle
  configurations of the related bispectrum, quantifying correlations between the
  galaxy overdensity field on large scales and the smaller-scale kSZ power.  We
  forecast the prospects for detecting this signal using future wide-field
  samples of Lyman-break galaxies from the Roman Space Telescope and
  next-generation CMB surveys including the Simons Observatory, CMB-S4, and
  CMB-HD.  We find that a roughly 13$\sigma$ detection is possible for CMB-HD
  and Roman after summing over all $\ell$ modes. We discuss the possibilities
  for improving this approach and related statistics, with the aim of moving
  beyond simple detections to measure the scale and redshift dependence of the
  cross-correlation signals.
\end{abstract}


\setlength{\footnotemargin}{\parindent}

\section{Introduction}
\label{sec:intro}

The Epoch of Reionization (EoR) is the time period when early star-forming
galaxies and accreting black holes first formed, emitted ultraviolet (UV) light,
and gradually photoionized hydrogen throughout essentially the entire volume of
the universe. There has been a great deal of recent observational progress in
studying the EoR
\citep{planck2018,Becker:2014oga,mcgreer_etal2015,davies_etal2018,mason_etal2019},
but future efforts are required to better understand when precisely reionization
occurs and to measure the spatial variations in the ionization state of the
intergalactic medium (IGM) during the reionization process. An improved
understanding here will reveal the properties of the universe during this
largely unexplored observational frontier, and help in testing models of the
earliest phases of galaxy and cosmological structure formation \citep{Loeb2013}.

One powerful signature of the EoR is referred to as the patchy kinetic
Sunayev--Zel'dovich (kSZ) effect
\citep{sunyaev_zeldovich1972,Gruzinov:1998un,Knox:1998fp}.  The ionization state
of the IGM during reionization is expected to be highly inhomogeneous, with
``bubbles'' of ionized gas forming around highly clustered ionizing sources,
which gradually grow and merge as the ionizing photons propagate into the IGM
and new UV-luminous sources turn on. Some of the photons from the cosmic
microwave background (CMB) scatter off of free electrons during the EoR and
receive Doppler shifts owing to the line-of-sight bulk peculiar velocities of
the ionized regions.  This imprints secondary anisotropies in the CMB, which are
sourced in part by the patchy ionized structure in the EoR. The CMB anisotropies
from the kSZ effect are hence sensitive to the bubble sizes during reionization
(e.g. \citealt{McQuinn:2005ce}).  Furthermore, the overall duration of
reionization partly sets the amplitude of the patchy kSZ signal
\citep{Gruzinov:1998un}.  In longer reionization scenarios, it is more likely
that CMB photons scatter off of free electrons within ionized bubbles during
reionization; this imprints larger-amplitude features on the CMB maps on the
angular scales spanned by the ionized regions.  Hence, the patchy kSZ signal
contains a wealth of information regarding the timing, duration, and spatial
structure of the reionization process
\citep{battaglia_etal2013b,natarajan_etal2013,choudhury_etal2021,paul_etal2021}.

There are, however, a number of challenges that must be overcome to extract this
signal from CMB surveys and to best exploit it as a probe of reionization. The
first issue is that one must separate the kSZ contributions to the CMB
anisotropies from other sources of variations at relevant frequencies, including
fluctuations in the cosmic infrared background (CIB) sourced by dusty
star-forming galaxies, radio sources, and the thermal Sunyaev--Zel'dovich (tSZ)
effect, predominantly from clusters and groups at much lower redshift. In
principle, these can be distinguished spectrally because the kSZ effect
maintains a blackbody spectral shape, in contrast to these other sources of
anisotropies. The kSZ fluctuations are, however, much smaller in amplitude and
so this places stringent requirements on the accuracy and precision of
component-separation algorithms
\citep{Zahn2012,Calabrese14,reichardt_etal2021}. This issue is further
complicated by spatial correlations between different components, such as the
tSZ and CIB, which must be properly accounted for \citep{Addison2012}.

A still more daunting challenge is separating the patchy reionization
anisotropies from post-reionization contributions to the kSZ effect, the
so-called ``late-time kSZ'' signal. The late-time kSZ effect shares, of course,
the blackbody spectral shape of the patchy component, and so must be
distinguished based on its angular dependence alone. Unfortunately, the expected
angular power spectrum of the post-reionization signal is relatively similar in
shape and amplitude to the contributions from the EoR \citep{Zahn2012}.
Furthermore, the late-time kSZ power spectrum itself is somewhat uncertain
because it depends, for instance, on feedback effects from galaxy formation and
supermassive black holes, which are hard to model \citep{park_etal2018}.
Finally, the kSZ signal is a projected quantity and so by itself provides mainly
an integral-type constraint on the reionization history and not detailed
information on how reionization evolves with redshift.  One promising approach
to help with these challenges is to exploit a suitable four-point function
statistic, which can aid in extracting the patchy reionization contributions and
their redshift evolution \citep{Smith:2016lnt,Ferraro:2018izc,Alvarez:2020gvl}.

Another possibility, which we pursue here, is to turn to cross-correlations. The
naive choice of a two-point correlation between the kSZ signal and another
tracer of large-scale structure (LSS) during the EoR is not, itself, promising:
In this statistic, a cancellation occurs because ionized regions may be moving
either toward or away from the observer.\footnote{On large scales
  ($\ell \sim 100$) this cancellation is avoided owing to rapid redshift
  evolution in the ionization fraction, but the expected correlation is weak and
  impractical to measure
  \citep{alvarez_etal2006,adshead_furlanetto2008,alvarez2016}.} This issue may
be easily circumvented, however, by simply filtering the CMB map to suppress
contamination from primary anisotropies and other contributions and then
squaring the filtered map
\citep{dore_etal2004,ferraro_etal2016,Hill:2016dta}. The filtered and squared
map will then correlate with a suitable LSS tracer, and photometric samples are
sufficient here, i.e. expensive spectroscopic observations are not
required. Indeed, this basic approach has been used successfully to study the
late-time kSZ effect by combining Planck CMB data with photometric galaxy
catalogs from the WISE survey at $z \lesssim 1.5$ to measure the kSZ$^2$--galaxy
cross-power spectrum \citep{Hill:2016dta,kusiak_etal2021}. Moreover, the future
outlook for more precise measurements of this statistic appears outstanding,
with forecasts promising measurements at many hundreds of sigma statistical
significance (\citealt{ferraro_etal2016}, hereafter
\citetalias{ferraro_etal2016}).  In the context of reionization studies,
\citet{ma_etal2018} studied the possibility of measuring the kSZ$^2$--21\,cm
cross-power spectrum, while \citet{laplante_etal2020} explored the related
kSZ-kSZ-21\,cm bispectrum. Both of these correlations with redshifted 21\,cm data
are, unfortunately, problematic in that the kSZ signal is sensitive only to
Fourier modes with very long-wavelength line-of-sight components, yet these
modes are almost inevitably lost to foreground contamination in the 21\,cm
surveys \citep{laplante_etal2020}.

Here we instead study, for the first time, the kSZ$^2$--galaxy cross-power
spectrum during the EoR.\footnote{\citet{Baxter:2020dvd} study the
  cross-correlation between the tSZ signal and reionization-era galaxy
  surveys. The tSZ fluctations from the EoR contain interesting information
  about the thermal state of the IGM during reionization but, as discussed in
  that work, are weaker than the kSZ anisotropies from reionization considered
  here.} This is motivated in part by the unprecedented photometric samples of
reionization-era galaxies expected from the Roman Space Telescope and
anticipated advances in CMB observations, with an upcoming suite of wide-field,
high-sensitivity and fine-angular-resolution surveys, including the Simons
Observatory (SO\footnote{\url{https://simonsobservatory.org}};
\citealt{so_science_goals}), CMB-S4\footnote{\url{https://cmb-s4.org}}
\citep{cmbs4_2017}, and CMB-HD\footnote{\url{https://cmb-hd.org}}
\citep{cmb_hd}.  In particular, the Roman high-latitude survey (HLS) will use
the Lyman-break selection technique to detect more than $\sim 10^6$ galaxies at
$z \geq 6$ across roughly 2200 deg$^2$ on the sky \citep{wfirst}. The
photometric galaxy samples provide clean measurements of transverse modes and
so, unlike the case of the foreground-corrupted 21\,cm field, are well suited
for combining with kSZ data. Furthermore, the high-redshift galaxy samples
should be well correlated with the patchy kSZ signal but not the late-time kSZ
effect. Likewise, most of the other variations at the relevant frequencies, such
as the tSZ, CIB, and radio sources, are produced at lower redshifts and so
should be uncorrelated -- or very weakly correlated -- on average with the
$z \geq 6$ Lyman-break galaxies from the Roman HLS sample. In other words, other
sources of anisotropy in the CMB maps will produce a noise source for
kSZ$^2$--galaxy correlation measurements but not an average bias. The exception
here is CMB lensing, which will contribute to our estimates and must be
separated based on its rather different angular dependence
\citepalias{ferraro_etal2016} (see also Section~\ref{sec:lensing_leakage}).

Finally, if the measurements can be made precisely enough, the kSZ$^2$--galaxy
cross-power spectrum may be measured in different photometric galaxy redshift
bins to study the evolution of the patchy kSZ effect with redshift. Thus, the
kSZ$^2$--galaxy cross-correlation offers a potential means to robustly extract
the reionization-era contributions to the kSZ signal and may also access
further tomographic information regarding the reionization history.

These features motivate the current work, which aims to model the kSZ$^2$--galaxy
cross-power spectrum and to forecast the expected signal-to-noise ratio (S/N)
achievable with upcoming surveys. In order to do so, we make use of seminumeric
simulations of reionization, which are fast and flexible, yet compare reasonably
well to more detailed radiative transfer calculations
\citep{battaglia_etal2013a}. In Sec~\ref{sec:methods}, we describe the
simulation methods used throughout this paper. In Section~\ref{sec:results}, we
present the simulated kSZ$^2$--galaxy cross-power spectrum models and discuss
their interpretation. In Section~\ref{sec:snr} we give S/N forecasts for a number
of upcoming surveys and describe the prospects for actually measuring the
simulated signals. Section~\ref{sec:future_improvements} discusses the
possibilities for more precise measurements further in the future. We conclude
and mention possible directions for extensions to this paper in
Section~\ref{sec:conclusion}. Throughout we assume a flat $\Lambda$CDM cosmology
with $\Omega_m = 0.316$, $\Omega_b = 0.049$, $h = 0.673$, $\sigma_8 = 0.812$,
and $n_s=0.966$. These parameters are consistent with those reported by the
Planck18 analysis \citep{planck2018}.

\section{Methods}
\label{sec:methods}

Here we present our basic approach for cross-correlating kSZ and galaxy survey
data. First, we describe the seminumeric reionization simulations used
throughout this work (Section~\ref{sec:zreion}). We then discuss how the mock
kSZ data (Section~\ref{sec:ksz}) and simulated maps of the galaxy distribution
are generated (Section~\ref{sec:galaxy}). Finally, we introduce the
kSZ$^2$--galaxy cross-power spectrum statistic employed in this paper
(Section~\ref{sec:xcorr}).

\subsection{Seminumeric Reionization}
\label{sec:zreion}

Here we make use of so-called seminumeric simulations of reionization. This
approach provides a fast and flexible treatment of the reionization
process. Given the relatively weak signal modeled here, it is helpful to have
rapid reionization calculations: This allows us to span large volumes, to
average over many independent simulation realizations, and to explore a few
different reionization models. Although more-detailed reionization simulations
incorporating radiative transfer and hydrodynamics have been performed
\citep{trac_cen2007,trac_etal2008,iliev_etal2014,ocvirk_etal2016,ocvirk_etal2020},
the box sizes in these calculations are too small to capture the large scales of
interest for our current study.

One such seminumeric method for modeling reionization is \texttt{zreion}
\citep{battaglia_etal2013a}. This technique has been applied to studies of the
kSZ \citep{battaglia_etal2013b,natarajan_etal2013,laplante_etal2020} and 21\,cm
signals \citep{laplante_etal2014,laplante_ntampaka2019}.  This model
approximates each simulated volume element as either completely ionized or
completely neutral.  In order to compute the ionization field $x_i(\bm{r},z)$
for each point and redshift in the simulation volume (where $x_i = 0$ represents
neutral gas and $x_i = 1$ is ionized), we begin by defining a local redshift of
the reionization field $z_\mathrm{re}(\bm{r})$. This field expresses the
redshift at which a particular cell in the simulation volume is first
reionized. We then define the fractional fluctuation in this field,
$\delta_z(\bm{r})$, as
\begin{equation}
\delta_z(\bm{r}) \equiv \frac{\qty[z_\mathrm{re}(\bm{r}) + 1] - \qty[\bar{z} + 1]}{\bar{z} + 1},
\label{eqn:deltaz}
\end{equation}
where $\bar{z}$ is a model parameter that defines the midpoint of
reionization. The reionization field $\delta_z$ is assumed to be a biased tracer
of the matter overdensity field on large scales. This relationship is
quantified by a bias parameter $b_{zm}(k)$:
\begin{equation}
b_{zm}^2(k) \equiv \frac{\ev{\delta^*_z \delta_z}_k}{\ev{\delta^*_m \delta_m}_k} = \frac{P_{zz}(k)}{P_{mm}(k)}.
\end{equation}
This bias is parameterized as a function of Fourier mode $k$ in the following
way:
\begin{equation}
b_{zm}(k) = \frac{b_0}{\qty(1 + \frac{k}{k_0})^\alpha}.
\label{eqn:bias}
\end{equation}
We use the value of $b_0 = 1/\delta_c = 0.593$, where $\delta_c$ is the critical
overdensity in the spherical collapse model. The reionization history is then
determined by the set of model parameters $\{\bar{z},\alpha,k_0\}$. The midpoint
of reionization is fixed by $\bar{z}$, defined in Equation~(\ref{eqn:deltaz}),
whereas $\alpha$ and $k_0$ influence the duration of reionization as described
in Equation~(\ref{eqn:bias}).  \citet{battaglia_etal2013a} demonstrate that the
\texttt{zreion} model compares favorably with full radiation hydrodynamic
simulations of reionization. In particular, for a suitable choice of model
parameters, it reproduces the average ionization history and the cell-by-cell
ionization field in the more-detailed simulation (and hence also summary
statistics such as the power spectrum of the ionization field).

For this study, our fiducial model adopts $\bar{z} = 8$, $\alpha = 0.2$, and
$k_0 = 0.9$ $h$\,Mpc$^{-1}$.\footnote{The values of $\bar{z}$, $\alpha$, and
  $k_0$ in \citet{battaglia_etal2013a} were calibrated from hydrodynamic
  simulations and are used as our ``early'' simulation parameters discussed
  below in Section~\ref{sec:alt_histories}.} These values produce an ionization
history that is relatively late and extended and consistent with measurements of
the electron-scattering optical depth, $\tau$, from Planck \citep{planck2018},
as well as observations of the Lyman-$\alpha$ forest toward high-redshift
quasars and studies of Lyman-$\alpha$ emitters
\citep{mcgreer_etal2015,davies_etal2018,mason_etal2019}.

For the current work, we simulate realizations of the matter density field using
second-order Lagrangian perturbation theory (2LPT). This method is sufficiently
accurate for the present study, in which we are interested in high,
reionization-era redshifts (i.e., $z \gtrsim 6$ or so), and mostly consider
large spatial scales \citep{scoccimarro_etal1998,Lidz:2006vj}. Our simulations
track 1024$^3$ particles in a cubic simulation volume with a comoving side
length of 2~$h^{-1}$ Gpc. In what follows, we restrict our analysis to multipole
moments that are well sampled and resolved in the simulations. Although our
simulations capture multipoles from roughly $\ell \sim 20$--19,000, we
conservatively consider only $100 \leq \ell \leq 10,000$ in our analyses.

To generate an ionization field, we first use 2LPT to evolve the particles to
the midpoint of reionization, $\bar{z}$. The particles are deposited onto a grid
with triangular-shaped clouds (TSC), and the resulting matter density field
$\delta_m(\bm{r})$ is tabulated. We then apply the bias factor in
Equation~(\ref{eqn:bias}) to the matter density field, yielding
$\delta_z(\bm{k})$. The local redshift of the reionization field,
$z_\mathrm{re}(\bm{r})$, follows after applying an inverse FFT to
$\delta_z(\bm{k})$ and using Equation~(\ref{eqn:deltaz}) to relate
$\delta_z(\bm{r})$ and $z_\mathrm{re}(\bm{r})$. The ionization field at a
particular redshift, $z_0$, is set to $1$ provided $z_\mathrm{re}(\bm{r})$ is
greater than $z_0$ (meaning that portion of the simulation volume was reionized
at an earlier time) and $0$ otherwise.

\subsection{The kSZ Field}
\label{sec:ksz}

The kSZ effect results from CMB photons that scatter off of free electrons that
are moving with some peculiar velocity relative to the observer,
$\bm{v} \cdot \vu{n}$, where $\vu{n}$ is a unit vector along the
line of sight. The resulting CMB temperature fluctuation is
\citep{sunyaev_zeldovich1972}
\begin{equation}
\frac{\Delta T (\vu{n})}{T_\mathrm{CMB}} =  -\int \dd{\chi} g(\chi) \bm{q} \cdot \vu{n} 
\label{eqn:ksz}
\end{equation}
Here $\chi$ is the comoving distance along the line of sight to the scattering
location and $\bm{q}$ is the local electron momentum field with
$\bm{q} = \bm{v}(1 + \delta_m)(1+ \delta_x)/c$. The patchy reionization effects
arise through the spatial fluctuations in the ionization fraction,
$1 + \delta_x = x_i/\ev{x_i}$. Here $x_i$ is the ionization fraction at the
scattering location, and $\ev{x_i}$ is the global average ionization
fraction. The quantity $g(\chi)$ is the visibility function, which describes the
probability that a CMB photon scatters between $\chi$ and $\chi - d\chi$,
without subsequent scattering along its path to the observer. This may be
written as \citep{alvarez2016}
\begin{equation}
g(\chi) = \pdv{\left[e^{-\tau({\chi})}\right]}{\chi} = e^{-\tau(\chi)}\sigma_T n_{e,0} \ev{x_i} (1 + z)^2,
\label{eqn:wksz}
\end{equation}
where $\sigma_T$ is the cross section for Thomson scattering and
$n_{e,0} = \qty[1 - (4 - N_\mathrm{He})Y/4]\Omega_b\rho_\mathrm{crit}/m_p$ is
the mean electron number density at $z=0$.\footnote{Note that the density and
  ionization fraction fluctuations are included in the local electron momentum
  field, $\bm{q}$ (Equation~\ref{eqn:ksz}), and so $g(\chi)$ describes the
  scattering probability as a CMB photon propagates through the universe at the
  cosmic mean density and ionization fraction.} Here, $\tau(\chi)$ is the
electron-scattering optical depth between the observer and a comoving distance
$\chi$. In this equation, we can safely compute $e^{-\tau(\chi)}$ using the
global average reionization history because the electron-scattering probability
is small, and its variations across different lines of sight are unimportant
here.  We set the number of helium ionizations per hydrogen atom,
$N_\mathrm{He} = 1$, so that helium is singly ionized along with hydrogen. (We
can safely ignore doubly ionized helium because helium becomes twice ionized
mainly at significantly lower redshifts; \citealt{laplante_etal2017}.) The
contribution of free electrons from singly ionized helium leads to a weak
dependence on the helium mass fraction, $Y$. We use Equations~(\ref{eqn:ksz})
and (\ref{eqn:wksz}) to generate kSZ maps from our simulations.

The details of the kSZ map-generating procedure are as follows.  We first
construct the redshift of the reionization field $z_\mathrm{re}(\bm{r})$ as
described above in Section~\ref{sec:zreion}. We then build kSZ sight lines
through the volume, interpolating from the comoving coordinates in the
simulation box onto a fixed grid in angular coordinates $(\theta_x,
\theta_y)$. The matter density and velocity fields at each position along a
given line of sight are linearly interpolated from 2LPT simulation snapshots at
two nearby scale factors, $a$. The ionization field is subsequently determined
using the values of $z_\mathrm{re}(\bm{r})$. We then sum up the contributions
across the entire length of each sight line and produce a two-dimensional map of
the kSZ fluctuation using Equations~(\ref{eqn:ksz}) and (\ref{eqn:wksz}). Note
that we only include reionization-era contributions to the simulated kSZ signal
in this work because the late-time kSZ signal is only a source of noise in the
present study. We do, of course, account for the late-time kSZ signal in our
estimates of the variance of the cross-correlation signal (see
Section~\ref{sec:snr}).

\subsection{The Galaxy Field}
\label{sec:galaxy}

In the current study, we adopt a simple linear-biasing model for the spatial
distribution of the Lyman-break galaxies in the Roman HLS. We base our models on
the BlueTides simulation \citep{feng_etal2015,feng_etal2016}. This simulation
resolves reionization-era galaxies and has been used to make detailed
predictions for an HLS-like survey (\citealt{waters_etal2016}, hereafter
\citetalias{waters_etal2016}). We use the redshift-dependent linear-bias factor
from their Figure~14 and the number density of mock HLS Lyman-break galaxies
from their Figure~3 in what follows. In the future, it may be interesting to
move beyond the linear-biasing model adopted here (see,
e.g. \citealt{mirocha_etal2021} for one approach), but we expect the
linear-biasing description to be relatively accurate for the forecasts here
because most of the S/N comes from large angular scales (see
Section~\ref{sec:snr}).

In order to model the two-dimensional galaxy distribution as a function of
direction on the sky, $\vu{n}$, we generate sight lines through the simulation
box that are fixed in angular coordinates, $(\theta_x,\theta_y)$. The resulting
two-dimensional galaxy fluctuation field is
\begin{equation}
\delta_g(\vu{n}) = \int \dd{z} W_g(z) b_g(z) \delta_m\left[\chi(z) \vu{n},\chi(z)\right].
\end{equation}
Here $W_g(z)$ is a redshift window, which describes the probability that a mock HLS Lyman-break galaxy lies within a particular range in redshift for measurements in a given photometric redshift bin. We approximate this by a top-hat function in redshift, with a midpoint $z_0$ and a width $\Delta z$:
\begin{equation}
  W_g(z;z_0,\Delta z) = \begin{cases}
    W_0 & \abs{z - z_0} \leq \Delta z/2 \\
    0 & \mathrm{else}
  \end{cases},
  \label{eqn:wg}
\end{equation}
where the normalization $W_0$ is a constant chosen so that the window function
obeys the constraint that $\int \dd{z} W_g(z) = 1$.  The minimal redshift width,
$\Delta z$, which may be achieved in practice depends on the accuracy of the
redshift determination for the Lyman-break galaxies in the HLS. Several studies
\citep{bouwens_etal2015,finkelstein2016} find photometric redshift uncertainties
of $\sigma_z \sim 0.2$--0.3 for $z \gtrsim 6$ Lyman-break galaxies, as selected
using the color filters available for the Hubble Wide Field Camera 3 (WFC3). The
photometric redshift accuracy for Roman's Wide Field Instrument (WFI) may be
higher given its larger set of filters. In addition, a grism instrument on Roman
can provide spectroscopic redshifts for a subset of the galaxies in the HLS
sample. In this work, we will vary $\Delta z$ to test its impact on the results
but will generally consider redshift bins that are substantially coarser than
the photometric redshift uncertainties. In our S/N calculations, we ignore
contamination in the Lyman-break samples from lower-redshift interloping
galaxies and foreground stars (see, e.g., \citealt{finkelstein2016} for a
discussion.)

\begin{figure}[t]
  \centering
  \includegraphics[width=0.45\textwidth]{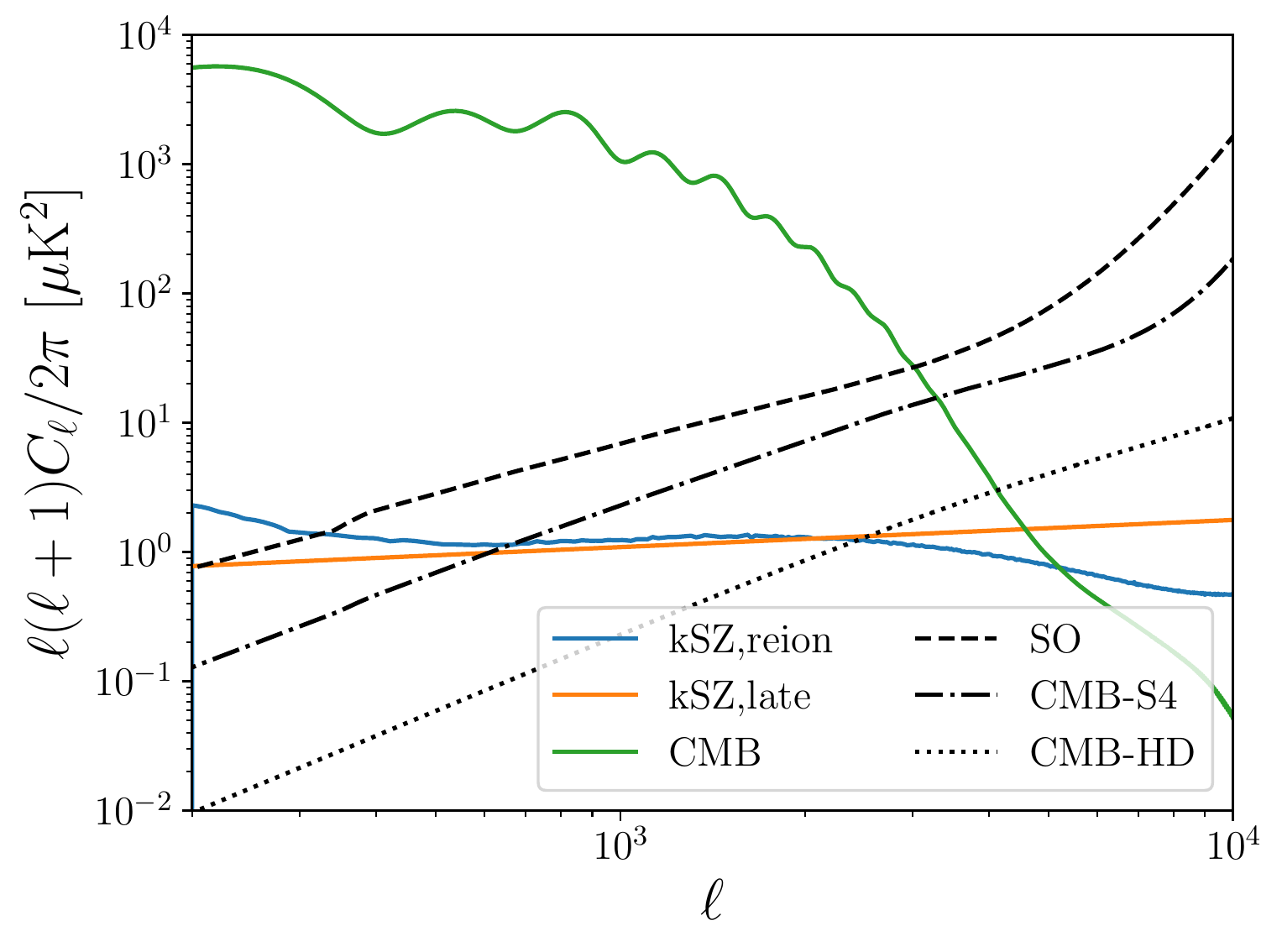}
  \caption{The various components of the filter defined in
    Equation~(\ref{eqn:wiener}). The kSZ signal from reionization is determined
    by averaging over 30 simulation realizations, as described in
    Section~\ref{sec:ksz}.  The late-time kSZ component is defined in
    Equation~(\ref{eqn:ksz_late}). The lensed primary CMB anisotropies are
    computed using CAMB \citep{lewis_etal2000}. The post-ILC noise power
    spectra, $N_\ell$, are shown for various CMB experiments.}
  \label{fig:cl_components}
\end{figure}

\subsection{Cross-correlation}
\label{sec:xcorr}

As motivated in the Introduction, we study the filtered kSZ$^2$--galaxy
cross-power spectrum in this work as a means to extract patchy reionization
contributions to the kSZ signal. The filter we apply suppresses $\ell$ modes
where the reionization-era kSZ angular power spectrum is small relative to other
contributions to the anisotropies on the relevant scales, which act as ``noise''
for the patchy kSZ measurement
\citep{dore_etal2004,ferraro_etal2016,ma_etal2018}. Specifically, the lensed
primary CMB anisotropies; residual foregrounds left over after component
separation from the CIB, radio sources, and the tSZ; the late-time kSZ signal;
and instrumental noise all contribute to the effective noise here. In this case,
an appropriate filter may be written as \citepalias{ferraro_etal2016}:
\begin{equation}
F(\ell) = \frac{C_\ell^{\mathrm{kSZ,reion}}}{C_\ell^{TT} + C_\ell^{\mathrm{kSZ,reion}} + C_\ell^{\mathrm{kSZ,late}} + N_\ell},
\label{eqn:wiener}
\end{equation}
where $C_\ell^{\mathrm{kSZ,reion}}$ is the angular power spectrum of the kSZ
signal due to reionization ($z \gtrsim 6$), $C_\ell^{TT}$ is the lensed primary
CMB spectrum, $C_\ell^{\mathrm{kSZ,late}}$ is the late-time ($z \lesssim 6$)
contribution to the kSZ, and $N_\ell$ accounts for both residual foregrounds and
detector noise for a given experiment, after deconvolving the instrumental
beam.  To generate $C_\ell^{\mathrm{kSZ,reion}}$, we run 30 independent
realizations of our simulations and average their $C_\ell$ spectra together. The
resulting maps have a power spectrum amplitude of
$\ell(\ell+1)C_\ell/2\pi \sim 1$ $\mu$K$^2$ at $\ell \sim 3000$, which is in
line with other seminumeric simulations of reionization for similar
reionization histories \citep{mesinger_etal2012,paul_etal2021}.  For the
late-time kSZ signal, we use the power-law fit of \citet{park_etal2018}, which
is calibrated from the Illustris simulations
\begin{equation}
D_\ell^{\mathrm{kSZ,late}} = 1.38 \qty(\frac{\ell}{3000})^{0.21}\,\mu\mathrm{K}^2,
\label{eqn:ksz_late}
\end{equation}
where $D_\ell = \ell(\ell+1)C_\ell/(2\pi)$.

\begin{deluxetable}{ccc}
  \tablecaption{Experiment Noise Properties \label{table:noises}}
  \tablehead{\colhead{Experiment} & \colhead{Map Noise ($\Delta_N$)$^1$} & \colhead{$\theta_\mathrm{FWHM}$}}
  \startdata
  Simons Observatory & 10 & $1.4'$ \\
  CMB-S4 & 2 & $1.4'$ \\
  CMB-HD & 0.6 & $10''$ \\
  \enddata
  \tablenotetext{1}{$\mu$K-arcmin}
  \tablecomments{These numbers define the noise values for the naive
    experimental noise defined in Equation~(\ref{eqn:beam}), as well as the beam
    size used in $b(\ell)$ (used both for the naive noise as well as the filter
    with the post-ILC noise estimates).}
\end{deluxetable}

In addition to this filter, we convolve the simulated signal maps with an
instrumental beam to account for the finite angular resolution of the upcoming
surveys. The beam is described by a Gaussian model and denoted by $b(\ell)$,
while the parameters for each instrument are listed in
Table~\ref{table:noises}. Accounting for both the noise-suppressing filter and
the instrumental beam, the observed temperature fluctuation $\Delta T_f(\ell)$
in harmonic space is related to the underlying variations before smoothing,
$\Delta T(\ell)$, as:
\begin{equation}
\Delta T_f (\ell) = F(\ell)b(\ell)\Delta T(\ell) \equiv f(\ell)\Delta T(\ell),
\label{eqn:filter_combined}
\end{equation}
where we have defined $f(\ell) \equiv F(\ell)b(\ell)$.

We consider the prospects for a range of upcoming CMB surveys, spanning the
noise ($N_\ell$) appropriate for each of SO, CMB-S4, and CMB-HD. Given the
frequency coverage of these surveys, residual foregrounds will inevitably remain
after component-separation algorithms use the blackbody spectral shape to
separate the kSZ signal from other contributions to the anisotropies. For
example, the standard internal linear combination (ILC) method constructs a
weighted linear combination of maps at different frequencies
\citep{bennett_etal1992,tegmark_etal2003,eriksen_etal2004,delabrouille_etal2009}. The
weights are chosen to give a unit response to the desired blackbody spectrum
and to minimize the variance of the resulting map. We consider applying the ILC
technique in harmonic, i.e., $\ell$ space, and use the publicly available
\texttt{orphics} code
\citep{hotinli_etal2021},\footnote{\url{https://github.com/msyriac/orphics}} to
estimate the expected angular power spectrum of residual CIB, radio sources,
tSZ, and instrumental noise. Note that the residual foreground contributions are
quite significant here, and so it is important to account for them along with
the instrumental noise.

\begin{figure}[t]
  \centering
  \includegraphics[width=0.45\textwidth]{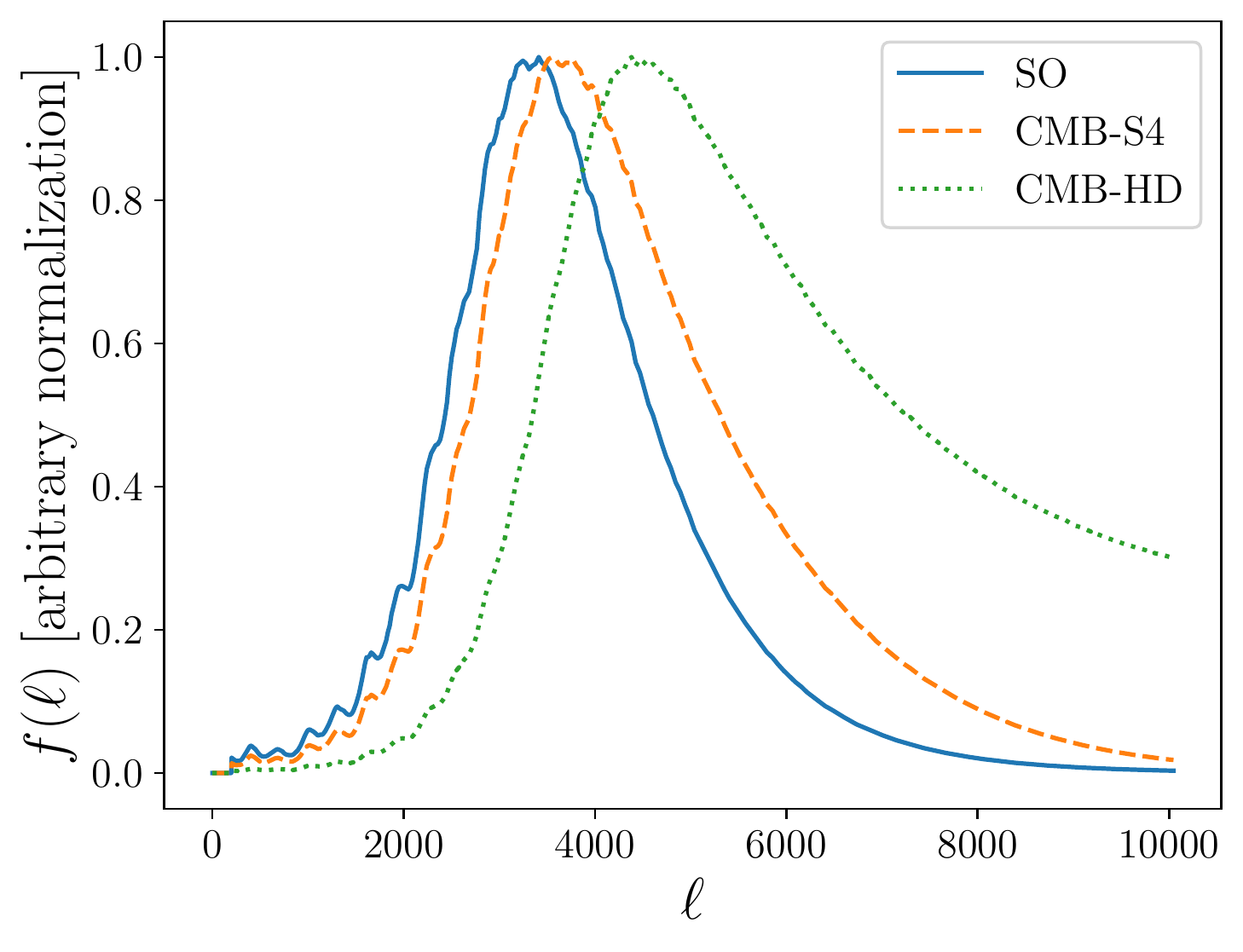}
  \caption{The combined filter and beam $f(\ell) = F(\ell)b(\ell)$ for the
    various experiments analyzed in this study.  The overall amplitude of this
    quantity is arbitrary, and so we normalize each curve to unity at the $\ell$
    value where the filter is maximal.  The filter downweights low $\ell$ modes
    where the primary anisotropies swamp the kSZ signal, as well as higher
    $\ell$ modes where residual foregrounds and instrumental noise dominate. In
    general, this filter peaks at higher $\ell$ values for the experiments with
    higher angular resolution and lower pixel noise. Note that although the
    CMB-HD filter would allow modes at $\ell \gtrsim 10,000$ such scales are not
    well resolved in our simulated maps but in any case contribute negligibly
    to the S/N. In practice, we hence truncate the filter beyond $\ell=10,000$.}
  \label{fig:fell}
\end{figure}

\begin{figure*}[t]
  \centering
  \mbox{}
  \includegraphics[height=135pt]{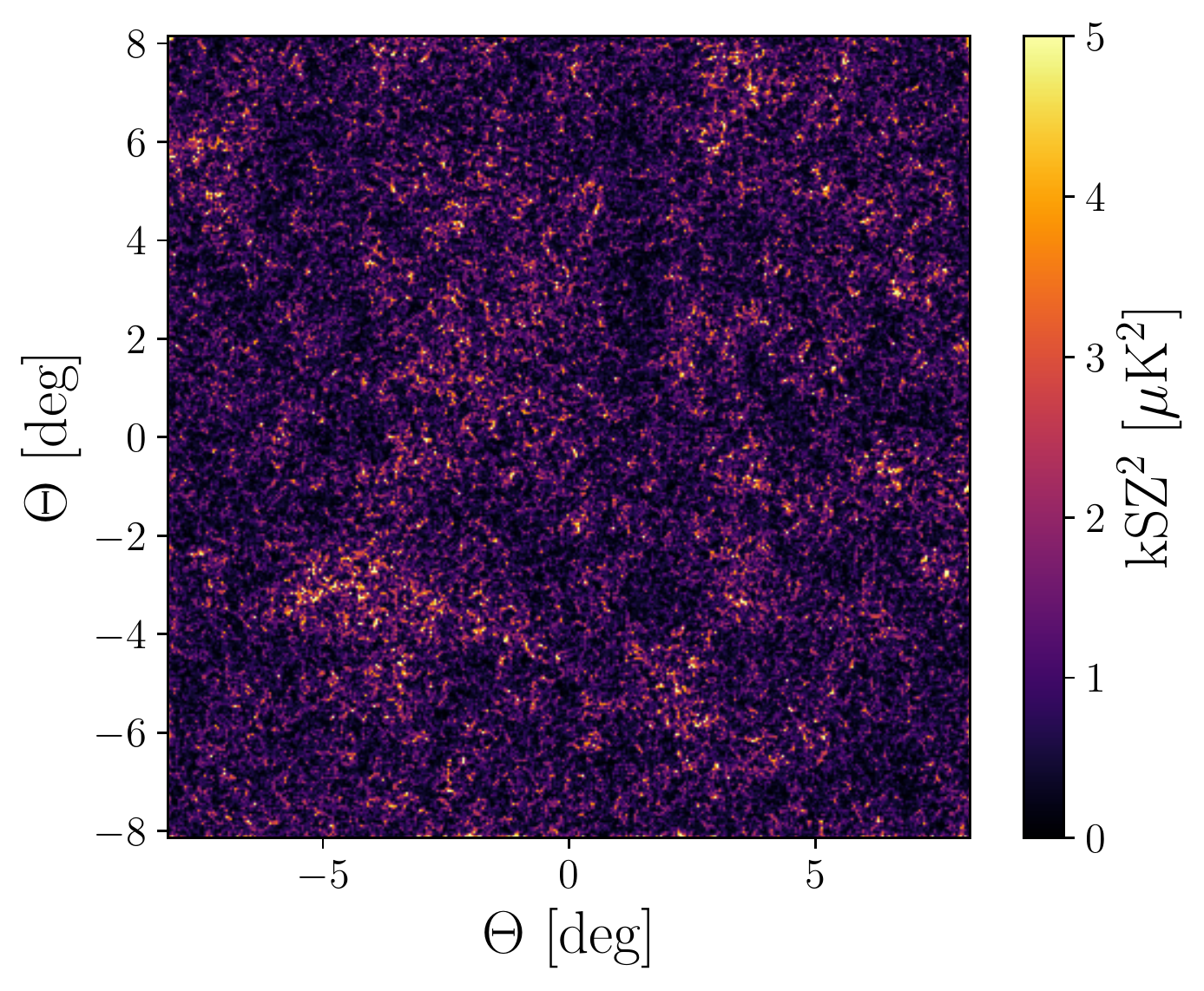}\hfill
  \includegraphics[height=135pt]{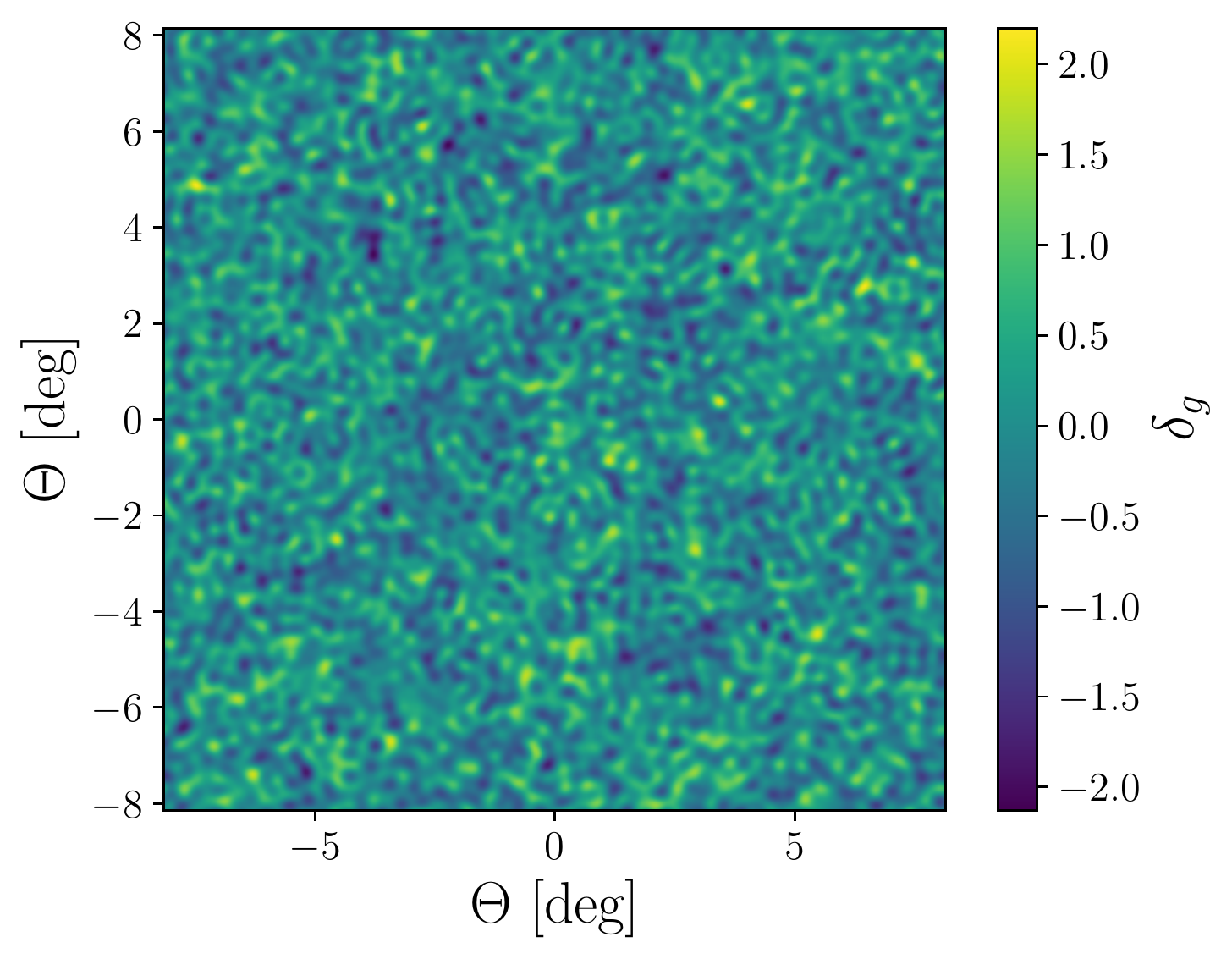}\hfill
  \includegraphics[height=135pt]{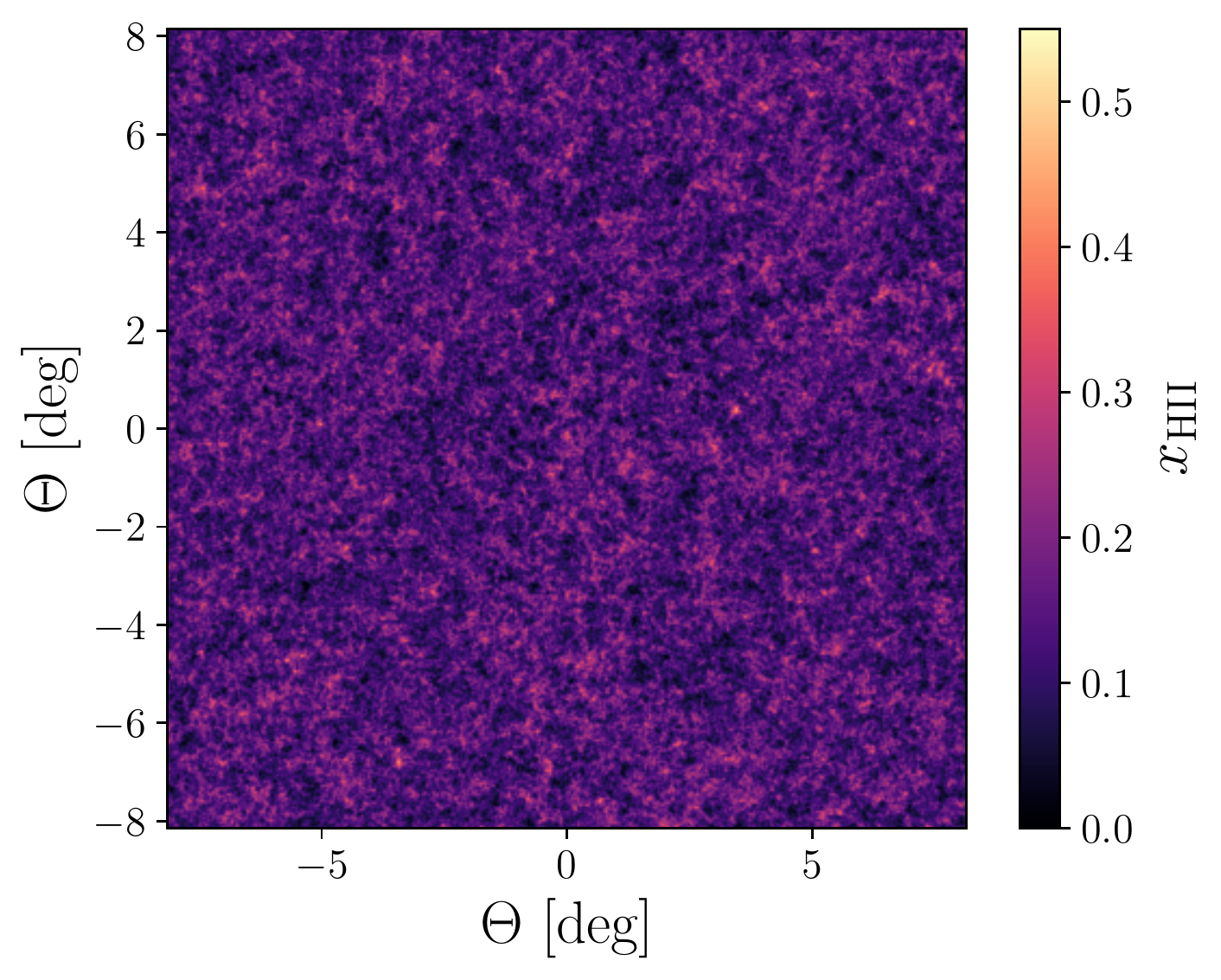}
  \mbox{}
  \caption{Left: a visualization of the kSZ$^2$ field from one of our
    simulations. Prior to squaring, we apply a filter
    $f(\ell) = F(\ell)b(\ell)$, with $F(\ell)$ defined in
    Equation~(\ref{eqn:wiener}) and $b(\ell)$ defined in
    Equation~(\ref{eqn:beam}). Here we use the parameters for SO.  Center: the
    corresponding high-redshift galaxy field. This two-dimensional field was
    generated by averaging across $9 \leq z \leq 10$, corresponding to a window
    function $W_g$ defined in Equation~(\ref{eqn:wg}) with $z_0 = 9.5$ and
    $\Delta z = 1$. We have also applied a low-pass filter removing all modes
    with $\ell \geq 1000$ to emphasize the large-scale features that are most
    important for the cross-correlation (discussed more below in
    Section~\ref{sec:results}). Right: the corresponding ionization field,
    $x_i$, averaged over $9 \leq z \leq 10$. The objective here is to use
    correlations between the kSZ$^2$ and galaxy fields to learn about the
    average ionization fraction evolution and its fluctuations.  Although the
    correlations are too weak to be seen by eye in the simulated maps here, they
    may be detected statistically by averaging over multiple simulation
    realizations, or by using larger patches of the sky in the real universe.}
  \label{fig:sim_maps}
\end{figure*}

For comparison, we will also consider cases where $N_\ell$ consists solely of
instrumental noise with negligible foreground contamination. In the
instrumental-noise-dominated limit, the relevant power spectrum is
\begin{equation}
N_\ell^\mathrm{naive} = \Delta_N^2 b^{-2}(\ell) \approx \Delta_N^2 \exp(\frac{\theta_\mathrm{FWHM}^2 \ell^2}{8 \ln 2}).
\label{eqn:beam}
\end{equation}
Table~\ref{table:noises} shows the various $\Delta_N$ and $\theta_\mathrm{FWHM}$
values assumed for each of the experiments \citep{hotinli_etal2021}. The more
realistic case, including residual foregrounds after ILC component separation
will be referred to as the ``post-ILC noise'' scenario, or the ``ILC noise''
case throughout.

Figure~\ref{fig:cl_components} shows the different components that contribute to
the filter defined in Equation~(\ref{eqn:wiener}). For multipoles of
$\ell \lesssim 4000$, the lensed primary CMB anisotropies are the dominant
source of noise for all upcoming CMB surveys, and significantly lower $\ell$
values are completely swamped by the primary anisotropies. Hence such scales
will be heavily downweighted in the filter. At higher $\ell$ values, residual
foregrounds and instrumental noise are the strongest contributions. While the
combined noise is larger than the patchy kSZ signal at all scales, it can
nevertheless be detected statistically after averaging over many $\ell$ modes.

The resulting filters for SO, CMB-S4, and CMB-HD are shown in
Figure~\ref{fig:fell}. As expected from Figure~\ref{fig:cl_components}, the
filter downweights low-$\ell$ modes that are swamped by the primary
anisotropies, as well as higher $\ell$ values where the residual foreground and
instrumental noise dominate.  In each case, the filter peaks at relatively
small scales near $3000 \lesssim \ell \lesssim 5000$, with the higher
resolution and more sensitive instruments giving peaks at smaller angular
scales.

\section{Results}
\label{sec:results}

We now turn to apply the filter of Equation~(\ref{eqn:filter_combined}) to the
simulated kSZ map, square the signal, and cross-correlate with the projected
galaxy distribution. It is instructive to first visually examine the resulting
simulated fields. Specifically, Figure~\ref{fig:sim_maps} shows the simulated
filtered kSZ$^2$ maps, generated following Section~\ref{sec:ksz}, along with
projected galaxy fluctuation fields (Section~\ref{sec:galaxy}) and ionization
fraction variations.  It is impossible to visually discern correlations between
the kSZ$^2$ signal and the ionization fluctuations because the kSZ signal
receives contributions from the entire reionization history and is sourced
partly by density and velocity variations. Likewise, the kSZ$^2$--galaxy
cross-correlation is too weak to see but may nevertheless be measured
statistically.  We include maps of the simulated signals just to provide some
overall impression of the quantities we model and correlate here.

The color bars in Figure~\ref{fig:sim_maps} are also instructive as they
indicate the amplitude of the fluctuations in these fields. For instance, the
very brightest fluctuations in the kSZ$^2$ map are at the kSZ$^2$
$\sim 30 \mu {\rm K}^2$ level (the color bar in Figure~\ref{fig:sim_maps} is
truncated above 5 $\mu$K$^2$ for visualization purposes), while the rms
fluctuation in this quantity is
$\mathrm{kSZ}^2_\mathrm{RMS} = 1.6$~$\mu\mathrm{K}^2$. Note that this is a
strongly non-Gaussian field, with a long large-fluctuation tail. The galaxy
abundance fluctuations are order $\sim$ unity, with an RMS of
$\delta_{g,\mathrm{RMS}} = 0.54$. Note that regions with $\delta_g < -1$ are
unphysical, and their occurrence here is an artifact of our linear-biasing
model. We nevertheless stick to this simple model in this work, as it should
still provide a reliable forecast for the expected S/N of the cross-spectrum. As
we will see, the kSZ$^2$ and galaxy fields are only weakly correlated and so the
cross-correlation signal is substantially smaller than the product of the RMS
variations in each field.

\begin{figure*}[t]
  \centering
  \mbox{}
  \includegraphics[width=0.32\textwidth]{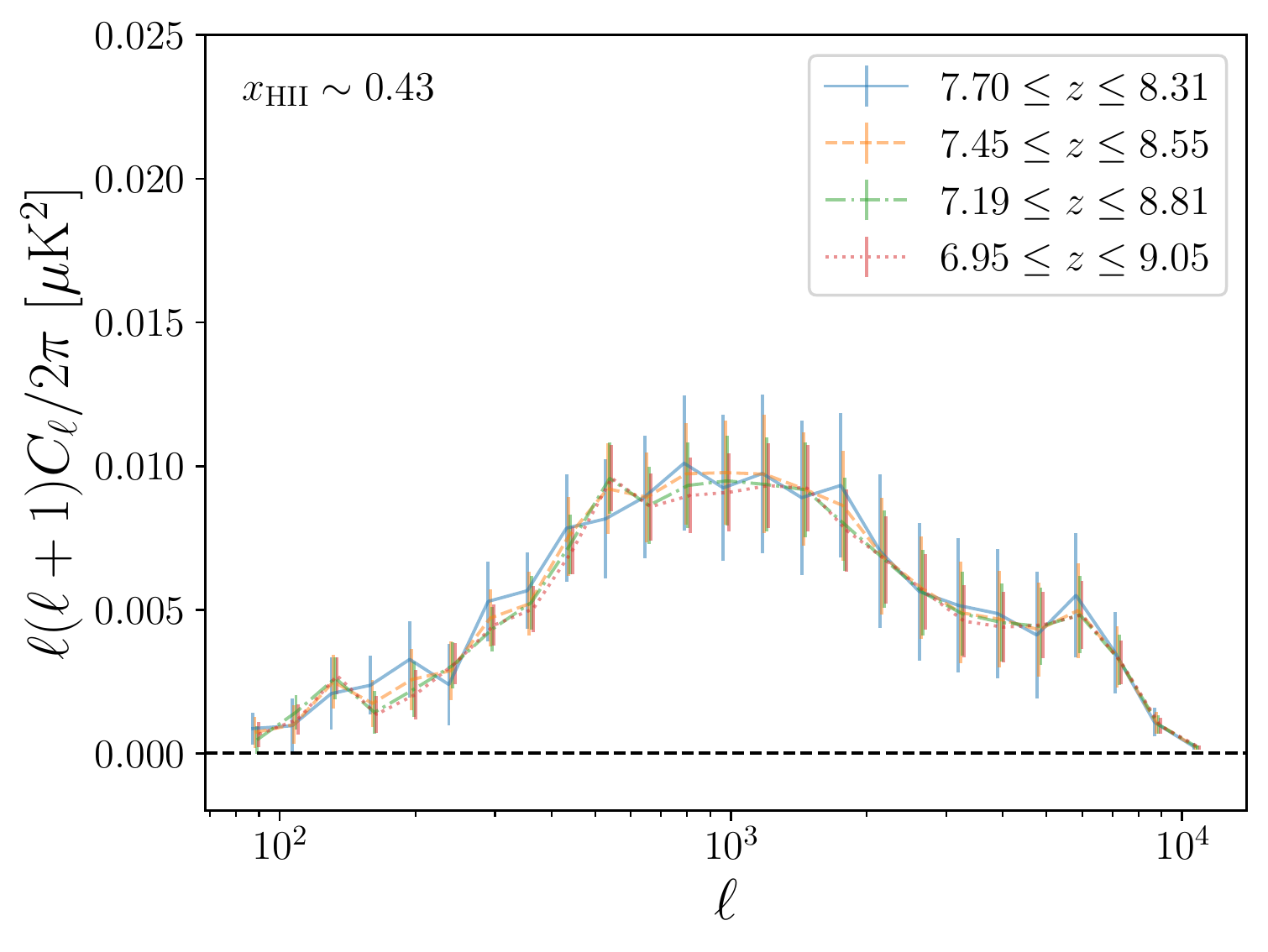}\hfill
  \includegraphics[width=0.32\textwidth]{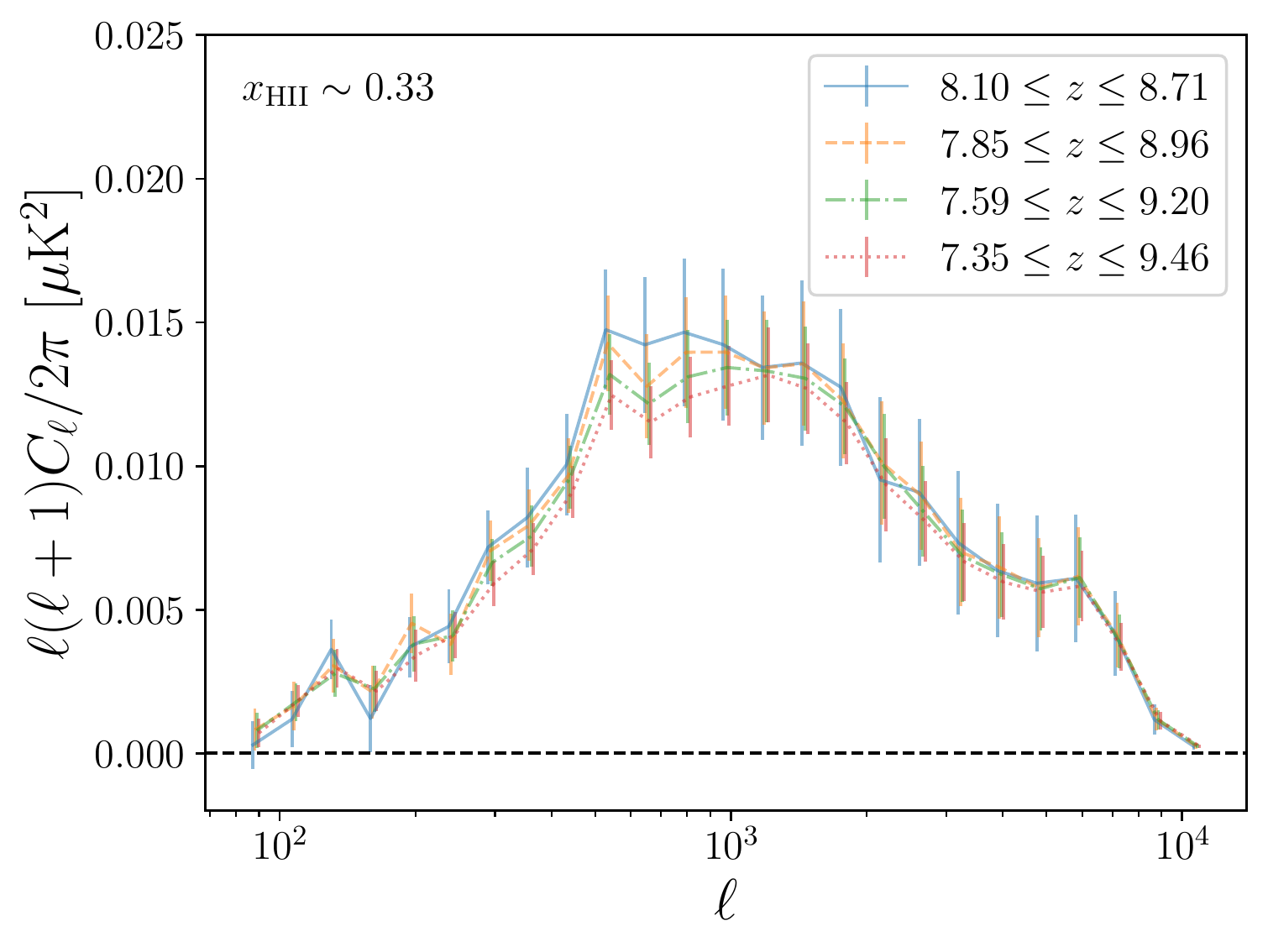}\hfill
  \includegraphics[width=0.32\textwidth]{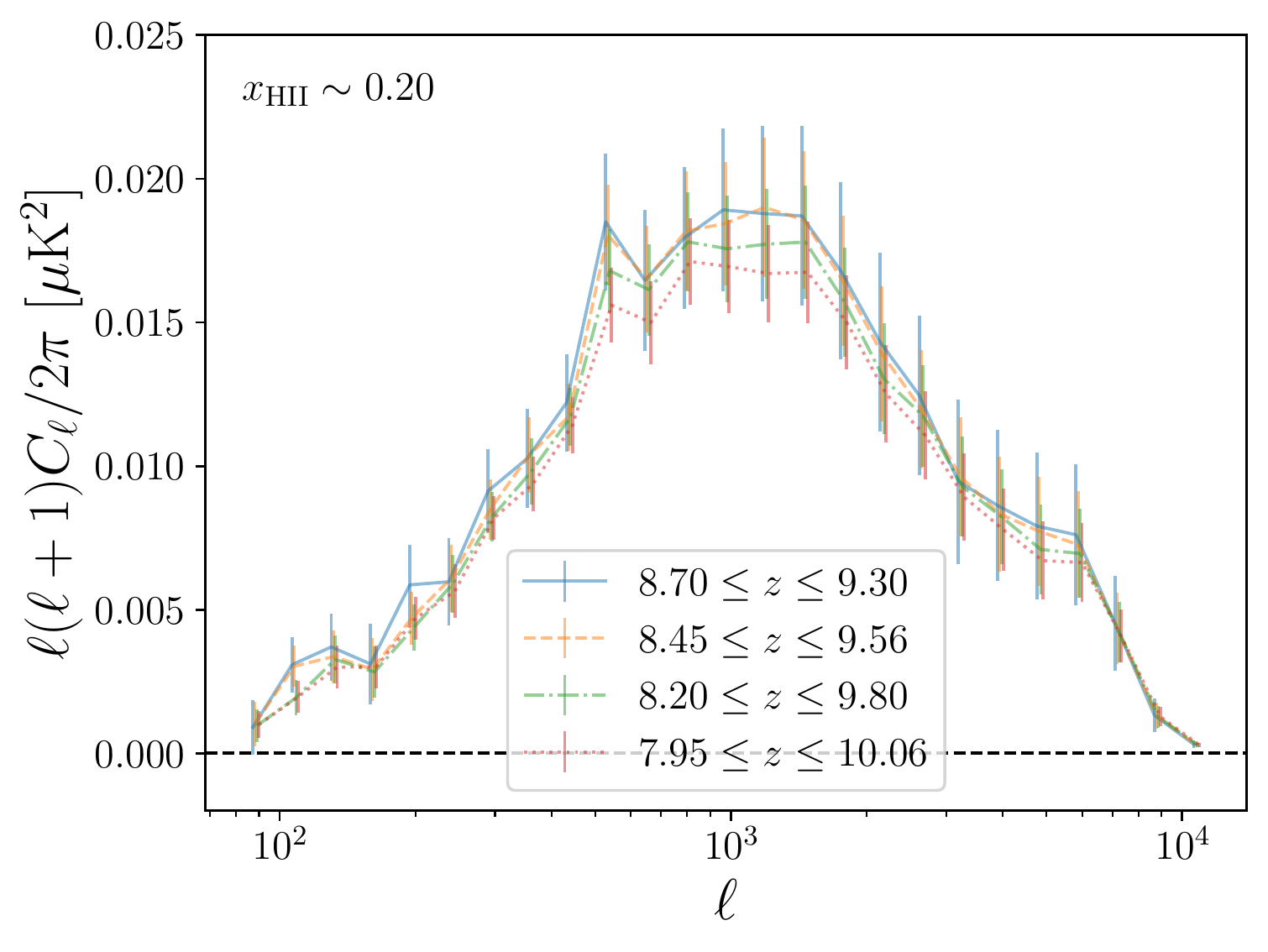}
  \mbox{}
  \caption{The kSZ$^2$--galaxy cross-correlation signal as a function of $\ell$ at different
    redshifts. Left: the signal for $z_0 = 8$,
    which corresponds to a volume-averaged ionization fraction of $x_{\mathrm{HII}} =
    0.43$. Different line colors correspond to different widths ($\Delta z$) of the galaxy redshift window. 
    The full redshift extent of the window is shown in the caption
    labels. Note that the lines have been slightly offset from each other in the
    horizontal direction for visual clarity. The error bars reflect the uncertainty
    on the simulation predictions, after averaging over $30$ independent simulation realizations.
    Center: windows centered on $z_0 = 8.4$, where
    $x_{\mathrm{HII}} = 0.33$. Right: windows centered on $z_0 = 9$, where
    $x_{\mathrm{HII}} = 0.2$. Further discussion can be found in
    Section~\ref{sec:results}.}
  \label{fig:kxg}
\end{figure*}

In order to quantify this, Figure~\ref{fig:kxg} shows the angular cross-power
spectrum between the kSZ$^2$ and galaxy fluctuation fields,
$C_\ell^{\mathrm{kSZ}^2\times\delta_g}$, as a function of $\ell$. Here we show
the results for the SO filter, although the results are broadly similar for the
CMB-S4 and CMB-HD filters, and consider various galaxy redshift distributions.
Specifically, we vary the central redshift, $z_0$, of the galaxy window, $W_g$
(Equation~\ref{eqn:wg}), and the redshift extent, $\Delta z$.  From
left to right, the central redshift values are $z_0 = \{8, 8.4, 9\}$, which
correspond to volume-averaged ionization fractions of
$x_{\mathrm{HII}} = \{0.43, 0.33, 0.2\}$. In order to reduce the noise on the
simulation measurements, we average over 30 independent simulation
realizations. The error bars in Figure~\ref{fig:kxg} give estimates of the error
on the mean simulated signal.

These results illustrate several interesting features of the signal. First, the
kSZ$^2$--galaxy cross-power spectrum peaks near $\ell \sim 1000$, virtually
independently of redshift. As discussed further below (see
Equation~\ref{eqn:bspec}), this implies that this statistic is mostly picking
out fairly squeezed-triangle configurations of the related bispectrum.  The SO
filter peaks at $\ell \sim 3000$, while the cross-power here is maximum near
$\ell \sim 1000$. The CMB-S4 and CMB-HD filters pick out still higher $\ell$
modes and the signal extends down to $\ell \sim$ a few hundred, so the statistic
mainly quantifies how the small-scale kSZ power varies with the galaxy
overdensity on larger scales. Although the signal is also appreciable at
$\ell \sim 3000$, corresponding to equilateral-type triangles, we will see that
most of the S/N comes from squeezed configurations (see Section~\ref{sec:snr}).
Second, the cross-correlation is positive for all redshifts and scales
shown. This arises because large-scale overdensities in the matter field contain
more galaxies, and also larger fluctuations in the ionization fraction on small
scales during most of the EoR, and hence greater high-$\ell$ kSZ power. Third,
the amplitude of the power spectrum peaks at around
$\ell (\ell +1) C_\ell^{\mathrm{kSZ}^2\times\delta_g}/(2 \pi) \sim
0.02$~$\mu$K$^2$. Although this is a small signal, one should keep in mind that
the filter applied suppresses the primary anisotropies and other sources of
variations in the map.  The peak is reached rather early in reionization: The
largest signal in the examples shown occurs in the rightmost panel, at a
volume-averaged ionization fraction of $x_{\rm HII} \sim 0.20$.  As discussed
further below, this results in part because the filter picks out kSZ variations
on fairly small angular scales.

Finally, another interesting feature of the results in Figure~\ref{fig:kxg} is
that the signal depends little on the redshift extent of the galaxy distribution
considered. We suspect that this reflects the following trade-offs. First, the
patchy kSZ signal receives contributions from a broad range of redshifts across
the EoR (Equations~\ref{eqn:ksz} and \ref{eqn:wksz}), and so increasing the
extent of the galaxy window boosts the overlap with the kSZ visibility
function. Second, however, the projected galaxy fluctuations decrease in
amplitude for a broader redshift window. Third, the CMB filter extracts the kSZ
fluctuations over a fairly narrow range of $\ell$ modes, concentrated on small
scales that peak relatively early in the EoR. Hence, although a broader galaxy
window promotes overlap with the visibility function, this does not boost the
signal appreciably because the small-scale kSZ$^2$ signal -- and its correlation
with matter overdensities -- is sharply peaked in redshift. It seems that, in
total, these three effects conspire to give little dependence on $\Delta z$.  We
discuss the interpretation of the simulated signal further below in
Section~\ref{sec:alt_histories}, but leave a more complete treatment to possible
future work.

Figure~\ref{fig:kxg_of_z} shows the redshift evolution of the cross-spectrum at
several example values of $\ell$ in more detail, with the corresponding
volume-averaged ionization fractions given along the upper $x$-axis. Here we
take galaxy redshift windows with a redshift extent of $\Delta z=1$ --
comfortably broader than the photometric redshift accuracy of the Lyman-break
galaxies detectable in the Roman HLS -- and show the central redshift in each
bin. Similar to Figure~\ref{fig:kxg}, the shaded regions show the standard error
of the mean computed from our 30 simulation realizations. The example $\ell$
modes in the figure all illustrate a similar trend with redshift and ionization
fraction: The signal rises and falls and peaks near an ionization fraction of
$x_{\rm HII} \sim 0.2$. The ionization fraction variations are evidently the
dominant source of signal here, as the cross-power is nearly vanishing towards
the end of reionization, around $z \sim 6$ in this model. This figure also
highlights the potential power of future kSZ$^2$-galaxy cross-power spectrum
measurements as a tomographic reionization probe. If the S/N of the measurements
is large enough, one can determine the cross-power spectrum in different
photometric redshift bins and extract how reionization evolves with
redshift. The small, yet nonvanishing, value of the cross-correlation in the
highest-redshift bin in Figure~\ref{fig:kxg_of_z} reflects contributions from
the low-redshift end of the bin. In this bin, a few percent of the simulation
volume is ionized, and the ionized regions are strongly concentrated around rare
overdense peaks in the matter distribution.

\subsection{Dependence on Ionization History}
\label{sec:alt_histories}

\begin{figure}[t]
  \centering
  \includegraphics[width=0.45\textwidth]{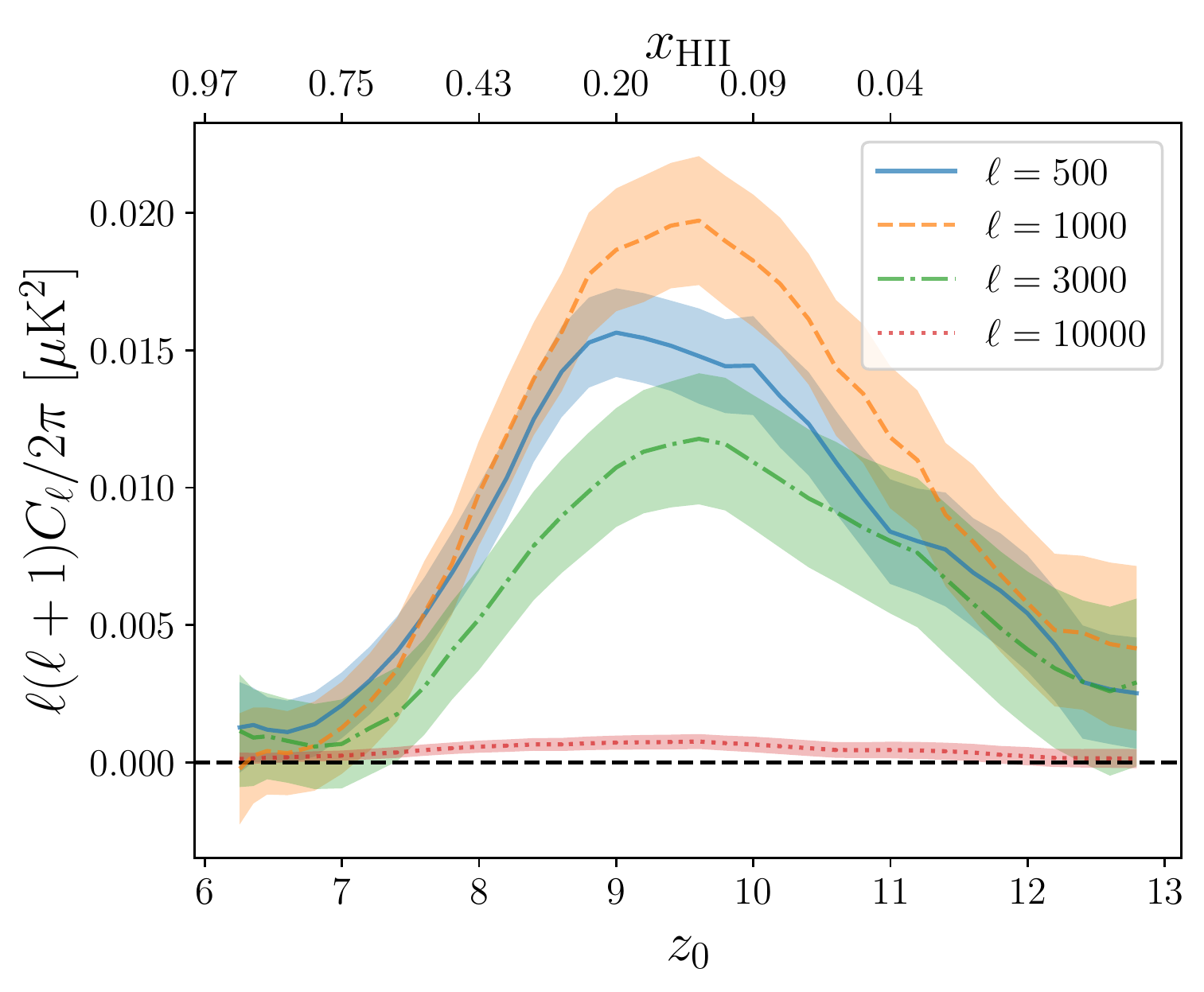}
  \caption{The cross-correlation statistic
    $C_\ell^{\mathrm{kSZ}^2\times\delta_g}$ as a function of $z_0$. Different
    lines correspond to different $\ell$ modes. The redshift values shown on the
    bottom $x$-axis correspond to the central redshifts, $z_0$, of the galaxy
    window, $W_g$. In each case the width is fixed to $\Delta z=1$. The
    corresponding volume-averaged ionization fractions, $x_{\rm HII}$, are shown
    on the upper $x$-axis.  For most modes, the signal peaks relatively early in
    reionization, near when the volume-averaged ionization fraction is
    $x_{\mathrm{HII}} \sim 0.2$.  The shaded regions show model uncertainties as
    in Figure~\ref{fig:kxg}. See Section~\ref{sec:results} for more discussion.}
  \label{fig:kxg_of_z}
\end{figure}

\begin{figure}[t]
  \centering
  \includegraphics[width=0.45\textwidth]{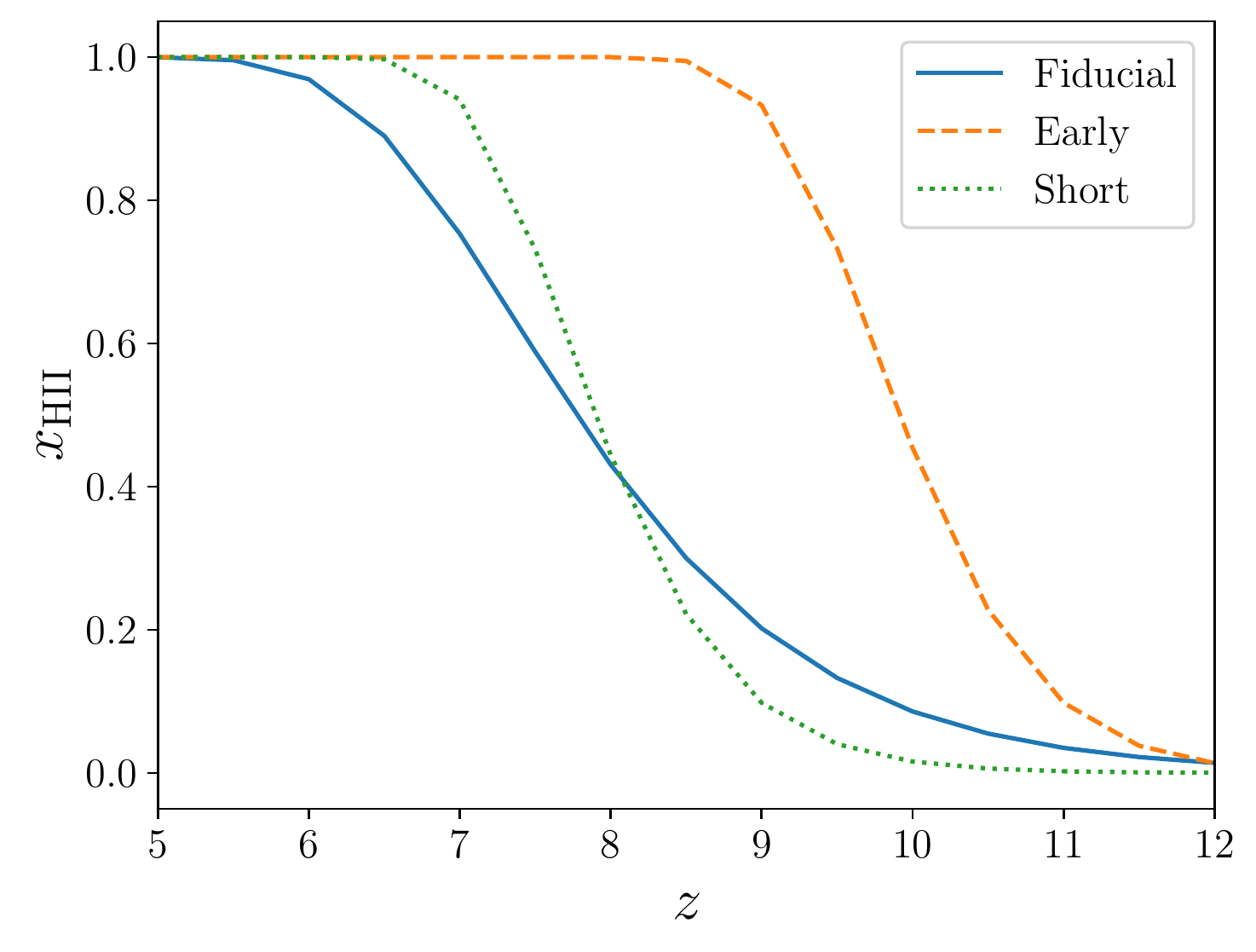}
  \caption{The volume-averaged ionization fraction of hydrogen,
    $x_\mathrm{HII}$, as a function of redshift for our fiducial, early, and
    short reionization scenarios. The different models shown here help in
    understanding how the kSZ$^2$--galaxy cross-power spectrum signal depends on
    the underlying reionization history.  }
  \label{fig:xhist}
\end{figure}

In order to explore further the trends with ionization fraction seen in
Figure~\ref{fig:kxg_of_z}, we produced two additional reionization models. In
these added cases, the timing of reionization differs from that in the fiducial
scenario considered thus far. Specifically, our alternate models include a
``short'' reionization history with the same midpoint of $\bar{z}=8$ (defined by
where $x_{\rm HII} \sim 0.5$), but a shorter reionization duration. We also
model an ``early'' reionization scenario, which has a similar overall duration
to our fiducial model, yet with a midpoint of $z=10$. The corresponding $k_0$
and $\alpha$ parameters (see Equation~\ref{eqn:bias}) are $k_0 = 0.185$
$h$Mpc$^{-1}$, $\alpha = 0.564$ for the short history. The early scenario uses
the same $k_0$ and $\alpha$ parameters as the short scenario but adopts a higher
value of $\bar{z} = 10$. The early scenario is disfavored by current
constraints, such as the electron-scattering optical depth inferred from
Planck18: the early scenario has $\tau_\mathrm{early} = 0.079$, whereas the
current Planck constraint is $\tau_\mathrm{Planck18} = 0.054 \pm 0.007$
\citep{planck2018}. Thus, the optical depth in the early scenario is 3.6$\sigma$
above the Planck18 central value.\footnote{The fiducial and short scenarios have
  $\tau_\mathrm{fiducial} = 0.058$ and $\tau_\mathrm{short} = 0.057$,
  respectively, and so are within 1$\sigma$ of the Planck18 measurement.}
Nevertheless, this scenario still provides a useful and illustrative example for
testing the impact of alternate reionization models on the kSZ$^2$--galaxy
cross-spectrum.  Figure~\ref{fig:xhist} contrasts the reionization histories in
the three different models explored here.\footnote{Note that we do not vary the
  abundance and clustering of the Roman HLS Lyman-break galaxies
  self-consistently with the reionization history model. That is, we just fix
  the bias factors and abundances of these sources to the BlueTides-model values
  throughout. This should be a good approximation in that the HLS observations
  will capture just the bright end of the galaxy populations and not the typical
  ionizing sources. Therefore, the observable Lyman-break galaxies are largely
  decoupled from the ionizing sources themselves.}

\begin{figure*}[t]
  \centering
  \includegraphics[width=0.45\textwidth]{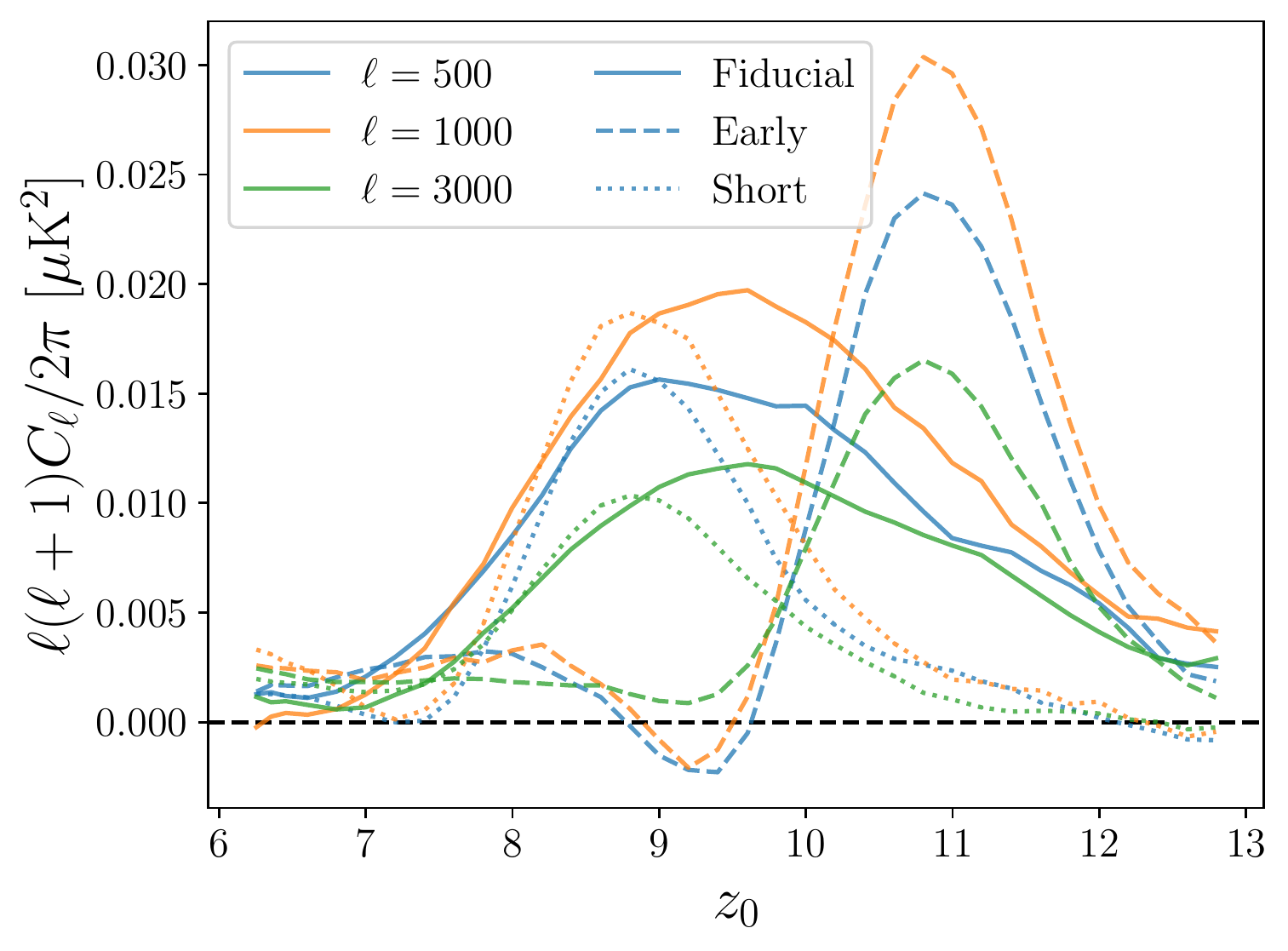}\hfill
  \includegraphics[width=0.45\textwidth]{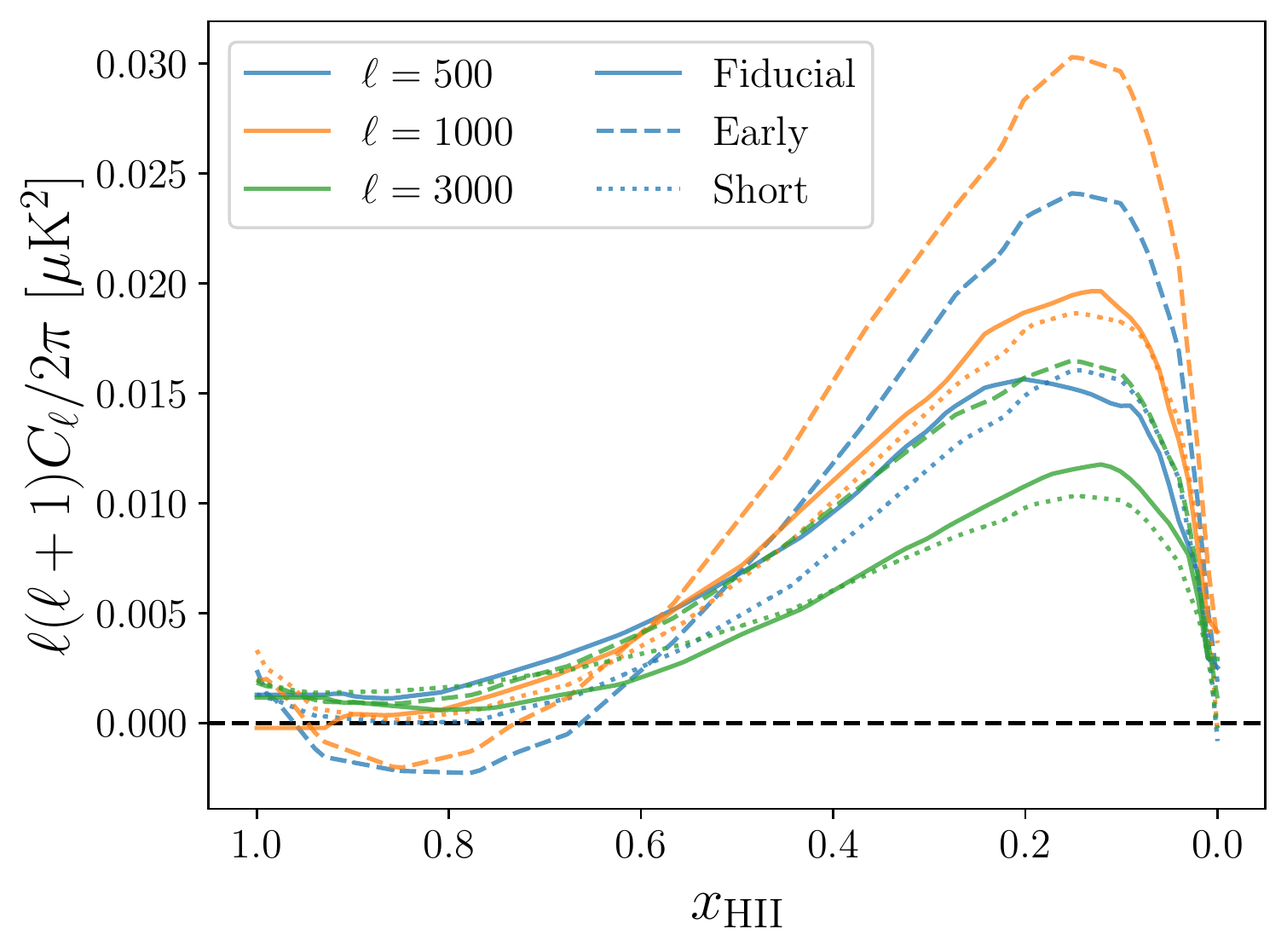}
  \caption{Left: the magnitude of $C_\ell^{\mathrm{kSZ}^2\times\delta_g}$ for
    various $\ell$ modes as a function of redshift, similar to
    Figure~\ref{fig:kxg_of_z}. This shows the behavior of the signal for the
    different ionization histories of Figure~\ref{fig:xhist}, discussed in
    Section~\ref{sec:alt_histories}. Right: the same as the left panel but
    plotted as a function of ionization fraction $x_\mathrm{HII}$ instead of
    redshift. Although the timing of reionization differs in each scenario, in
    all three cases the signal peaks at a similar average ionization
    fraction. The overall rise and fall in the amplitude of the cross-spectrum
    are also comparable when the three models are considered as a function of a
    ionization fraction. The generic behavior here is promising to use this
    statistic to extract information about the timing of reionization.}
  \label{fig:xhist_comp}
\end{figure*}

Figure~\ref{fig:xhist_comp} compares the behavior of the kSZ$^2$--galaxy
cross-power spectrum in the three different models. The left-hand panel shows
that the early model has the largest amplitude signal. This results because of
the increased gas density at high redshift, which leads to a larger
electron-scattering opacity at early times in this model. The increased opacity,
along with the enhanced clustering bias of the galaxies at high redshift, boosts
the signal in this scenario. Although the signal is larger, we will see that the
cross-spectrum is less detectable in this case (see Section~\ref{sec:snr}). The
fiducial and short reionization histories have similar peak amplitudes, but the
rise and fall with redshift in the fiducial model are more gradual than in the
short history. This illustrates that the cross-power spectrum statistic modeled
here can be used to determine how rapidly reionization evolves with redshift,
provided it can be measured precisely enough across different photometric
redshift bins.

Another interesting feature of Figure~\ref{fig:xhist_comp} is emphasized in the
right panel, in which the cross-spectrum is plotted as a function of the
volume-averaged ionization fraction $x_{\rm HII}$. Although the timing of
reionization differs markedly across our three example models, they all show a
similar evolution when considered as a function of $x_{\rm HII}$. In each case,
the signal rises rapidly at small ionization fractions, reaches a peak at around
$x_{\rm HII} \sim 0.15$--0.2, and then shows a more gradual decline with
increasing ionization fraction. One detail appears slightly differently in the
early reionization history, however: This model shows a negative
cross-correlation signal toward the end of reionization. This may result
because the large-scale matter densities are almost entirely ionized by the late
stages of reionization in this scenario. In this case, the ionization field is
smooth on small scales within overdense regions, and so there is less
small-scale kSZ power in such portions of the universe (leading to a negative
correlation). At any rate, the main result of Figure~\ref{fig:xhist_comp} is
that the kSZ$^2$--galaxy cross-power spectrum evolves in a generic way with
average ionization fraction across these three different scenarios. It may
therefore help to determine the timing of reionization, with the peak providing
a potential marker of when 15-20\% of the volume of the universe is ionized.

A more quantitative understanding of the trends observed in our simulated
examples may be obtained by relating the kSZ$^2$--galaxy cross-power spectrum to
an underlying bispectrum. We leave a more complete treatment along these lines
to possible future work and confine ourselves here to a few general
remarks. The cross-power spectrum may be written, in the Limber approximation,
as \citep{dore_etal2004}:
\begin{multline}
C_\ell^{\mathrm{kSZ}^2\times\delta_g} = \int \frac{\dd^2L}{(2 \pi)^2} f(L) f(\abs{\boldell + \mathbf{L}}) \times \\
 \int \frac{\dd{\chi}}{\chi^4} W_g(\chi) b_g(\chi) g^2(\chi) \mathcal{B}_{\delta_m p_{\hat{n}} p_{\hat{n}}}\left[\Big(\frac{\boldell}{\chi},\frac{\mathbf{L}}{\chi},\frac{\mathbf{-L}-\boldell}{\chi}\Big);z\right].
\label{eqn:bspec}
\end{multline}
The cross-bispectrum in the above equation,
$\mathcal{B}_{\delta p_{\hat{n}} p_{\hat{n}}}$, involves the matter density
fluctuations, $\delta_m$, and the spatial variations in the line-of-sight
component of the electron momentum field, $p_{\hat{n}}$ at two different
multipoles. Note that the ionization fraction fluctuations are embedded in the
electron momentum field here (see Equations~\ref{eqn:ksz} and \ref{eqn:wksz}).

A few immediate conclusions can be drawn from Equation~(\ref{eqn:bspec}). First,
recall that $f(L)$ is rather sharply peaked around
$3000 \lesssim L \lesssim 5000$, with the detailed filter shape depending on the
CMB survey considered (see Figure~\ref{fig:fell}). If we consider $\ell \ll L$,
the regime where the S/N peaks (see Section~\ref{sec:snr}), then the above
cross-spectrum is picking up contributions from the bispectrum for
squeezed-triangles configurations, as mentioned earlier. The squeezed-triangle
bispectrum describes the correlation between the low-$\ell$ fluctuations in the
galaxy density field and the high-$L$ kSZ power. Second, near the redshift of
the signal peak at $z \sim 9.5$, the CMB filter extracts primarily somewhat
small-scale electron momentum fluctuations with a wavenumber of
$k \sim L/\chi(z) = 0.3$--0.5 Mpc$^{-1}$ for $3000 \lesssim L \lesssim 5000$. At
these multipoles, the angular power spectrum of the density-weighted ionization
fraction fluctuations (in $\Delta z=1$ bins) peaks when the average ionization
fraction is $x_{\rm HII} = 0.38$. That is, the small-scale density-weighted
ionization power reaches its maximum fairly early in the reionization
process. This peak occurs, however, slightly after that in the kSZ$^2$--galaxy
cross-spectrum signal. This presumably relates to the fact that our statistic is
sensitive to how the small-scale kSZ variations correlate with the large-scale
galaxy fluctuations, and so its amplitude is not entirely set by the strength of
the small-scale electron momentum fluctuations alone.

In future work, it may be interesting to decompose the bispectrum of
Equation~(\ref{eqn:bspec}) into several constituent terms involving various
products of the ionization, density, and velocity fields. One can then carry out
the integrals in Equation~(\ref{eqn:bspec}) to quantify the relative
contributions of the different terms to the cross-spectrum and further analyze
their dependence on the galaxy window and reionization model.

\section{Observational Prospects}
\label{sec:snr}

We now turn to consider the prospects for detecting the kSZ$^2$--galaxy
cross-power spectrum with upcoming surveys. As motivated previously, we focus on
the case of the Roman HLS Lyman-break galaxy sample in combination with SO,
CMB-S4, and CMB-HD.  First, we start by determining the total S/N at which we
can measure $C_\ell^{\mathrm{kSZ}^2\times\delta_g}$ in a given photometric
redshift bin, summed over all $\ell$ modes. We consider various values for the
galaxy redshift window parameters, $z_0$ and $\Delta z$
(Equation~\ref{eqn:wg}). Here we follow the closely related calculations in
\citetalias{ferraro_etal2016}.

Assuming our fiducial case describes the true underlying kSZ$^2$-galaxy
cross-power spectrum signal, the total S/N of a measurement is
\citepalias{ferraro_etal2016}
\begin{equation}
\left(\frac{\rm S}{\rm N}\right)^2 
= f_\mathrm{sky} \sum_\ell \frac{(2\ell+1)\qty(C_\ell^{\mathrm{kSZ}^2\times\delta_g})^2}{C_\ell^{\bar{T}^2\bar{T}^2,f}C_\ell^{\delta_g\delta_g} + \qty(C_\ell^{\mathrm{kSZ}^2\times\delta_g})^2},
\label{eqn:snr}
\end{equation}
where $f_\mathrm{sky}$ is the fraction of overlapping sky surveyed by both
experiments, $C_\ell^{\delta_g\delta_g}$ is the angular power spectrum of the
galaxy map including shot noise, and $C_\ell^{\bar{T}^2\bar{T}^2,f}$ is the
total power spectrum of the filtered CMB$^2$ map. The equation above adopts the
usual Gaussian approximation for the cross-spectrum variance.  Furthermore, we
also compute $C_\ell^{\bar{T}^2\bar{T}^2,f}$ in the Gaussian approximation
\citepalias{ferraro_etal2016}:
\begin{equation}
C_\ell^{\bar{T}^2\bar{T}^2,f} \approx 2 \int \frac{\dd[2]{\bm{L}}}{(2\pi)^2} C_L^{\bar{T}\bar{T},f} C_{\abs{\bm{L}-{\boldsymbol\ell}}}^{\bar{T}\bar{T},f}.
\label{eqn:conv}
\end{equation}
Here we define
\begin{equation}
C_\ell^{\bar{T}\bar{T},f} \equiv f^2(\ell)\qty(C_\ell^{TT} + C_\ell^{\mathrm{kSZ,reion}} + C_\ell^{\mathrm{kSZ,late}} + N_\ell),
\label{eqn:ctt}
\end{equation}
where the various components have the same meaning as in
Equation~(\ref{eqn:wiener}).

\begin{figure}[t]
  \centering
  \includegraphics[width=0.45\textwidth]{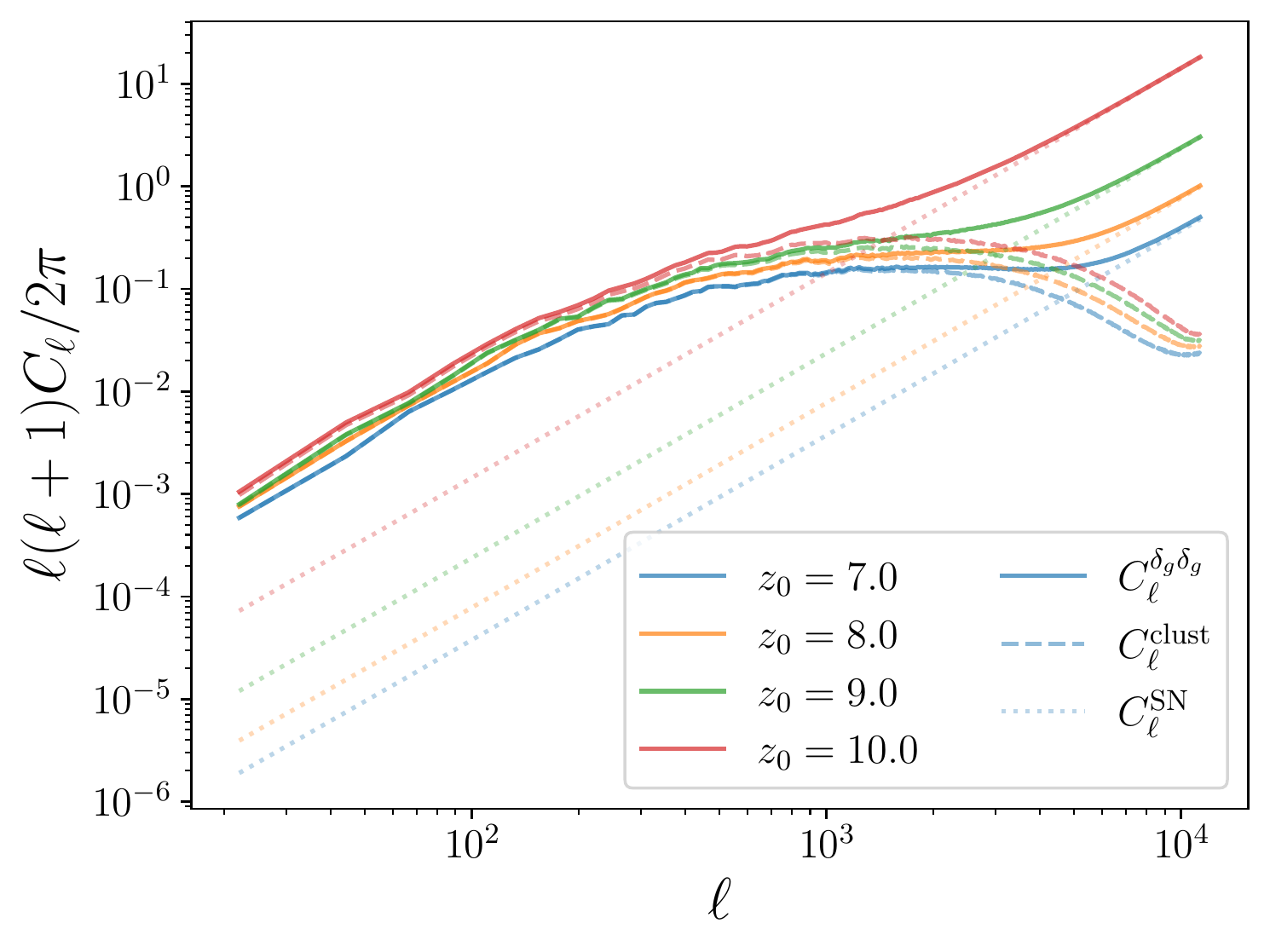}
  \caption{The galaxy power spectrum, as defined in
    Equation~(\ref{eqn:galaxy_cl}). The dashed lines show the clustering
    component, $C_\ell^{\mathrm{clust}}$, as estimated from our simulations. The
    central value of the galaxy selection window $z_0$ for each line is
    indicated in the legend, and the width is fixed to $\Delta z = 1$ in each
    case. The dotted lines give the shot-noise contribution
    $C_\ell^{\mathrm{SN}}$. The number of galaxies as a function of redshift and
    the linear bias factors are taken from \citetalias{waters_etal2016}. The
    solid line shows the sum of the two contributions,
    $C_\ell^{\delta_g\delta_g}$.}
  \label{fig:galaxy_cl}
\end{figure}

Equation~(\ref{eqn:snr}) requires an estimate of $C_\ell^{\delta_g \delta_g}$,
which includes both clustering and shot-noise terms. This is written as
\begin{equation}
C_\ell^{\delta_g\delta_g} = C_\ell^{\mathrm{clust}} + C_\ell^{\mathrm{SN}},
\label{eqn:galaxy_cl}
\end{equation}
where $C_\ell^{\mathrm{clust}}$ is the clustering component of the galaxy
$C_\ell$ spectrum estimated from our simulations and $C_\ell^{\mathrm{SN}}$ is
the shot-noise contribution. Both the clustering and shot-noise terms depend on
the galaxy window, $W_g$. As discussed previously, we assume a linear-biasing
model and use the galaxy bias versus redshift from \citetalias{waters_etal2016}.
The shot noise is computed as
$C_\ell^{\mathrm{SN}} = \frac{f_\mathrm{sky} 4\pi}{N_g},$ where $N_g$ is the
total expected number of galaxies for the HLS survey in a given photometric
redshift bin.  Throughout this analysis, we assume a survey area for the HLS of
2200 deg$^2$ \citep{wfirst}, which translates to $f_\mathrm{sky} = 0.053$. We
suppose here that the CMB survey covers the entire HLS field. We use results
from the BlueTides simulations in \citetalias{waters_etal2016} for the expected
number of HLS galaxies and the corresponding linear-bias factors.  Specifically,
we adopt their simulated galaxy abundance and clustering for the case of a
5$\sigma$ detection threshold. We take the geometric mean between the optimistic
``intrinsic'' brightness of the simulated galaxy samples and their
``dust-corrected'' galaxy abundance model. These values are broadly consistent
with the predicted number of galaxies from \citet{wfirst}, which are based off
of measurements from \citet{bradley_etal2012} at $z \sim 8$ and extrapolated to
higher redshifts, assuming a model similar to that of \citet{postman_etal2012}.

\begin{figure}[t]
  \centering
  \includegraphics[width=0.45\textwidth]{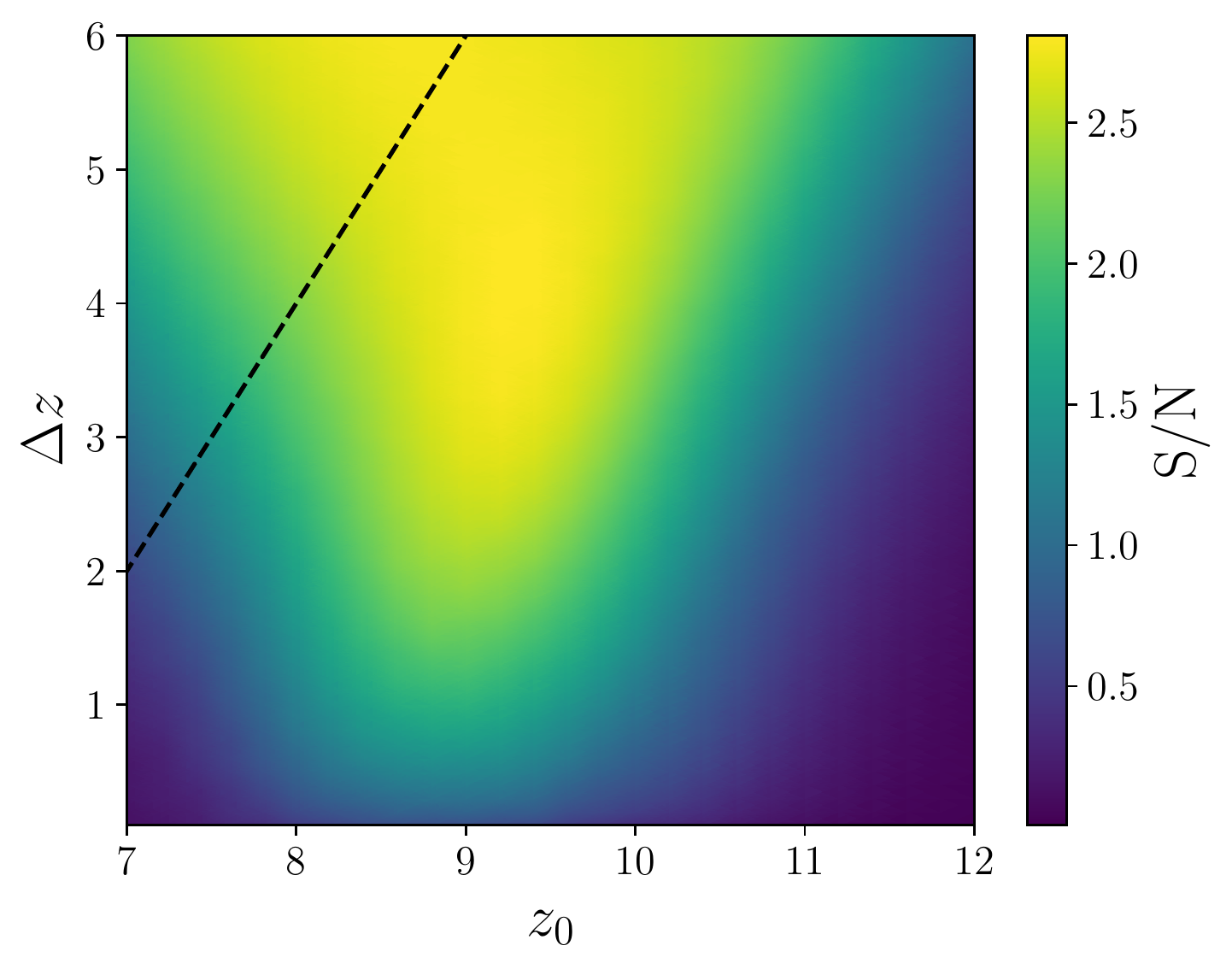}
  \caption{The S/N from Equation~(\ref{eqn:snr}), for a single redshift bin, as
    a function of the galaxy window function $W_g$ parameters $z_0$ and
    $\Delta z$. This shows the result for CMB-S4 assuming the fiducial
    reionization scenario. The maximum response comes from a relatively wide
    window of $\Delta z \sim 4.2$, centered near $z \sim 9.2$. Interestingly,
    the peak in S/N is at a slightly smaller redshift than the peak in the
    signal, owing primarily to the smaller galaxy shot-noise term at lower
    redshift. The corresponding S/N is 2.8$\sigma$. The dashed line in the
    upper-left part of the figure corresponds to combinations of $z_0$ and
    $\Delta z$ that would include contributions from post-reionization
    ($z \leq 6$), which are not included in our simulations. See
    Section~\ref{sec:snr} for further discussion.}
  \label{fig:snr}
\end{figure}

Figure ~\ref{fig:galaxy_cl} shows the angular power spectrum of the galaxy
distribution in our models as a function of redshift. An interesting feature
here is that, perhaps contrary to naive expectations, the Roman HLS galaxy
samples are not entirely swamped by shot noise, as emphasized earlier by
\citetalias{waters_etal2016}. For instance, in the case shown, which adopts a
photometric redshift bin of $\Delta z=1$, the shot noise is subdominant to the
clustering term for $\ell \lesssim 1000$ even at $z \sim 10$.  This is partly a
consequence of the linear-bias factors, which are expected to be huge for the
Roman HLS galaxies, ranging from $b_g \sim 9$ at $z=6$ to $b_g \sim 20$ at
$z=12$. Note also that the clustering term actually increases slightly toward
high redshift, even though the matter density fluctuations are smoother at early
times. This occurs because $b_g$ grows relatively steeply with redshift, while
at the same time the comoving volume enclosed within a shell of fixed redshift
width $\Delta z$ decreases. These effects tend to boost the fluctuations in the
projected galaxy distribution toward high redshift and more than compensate for
the smoother matter field. Nevertheless, as can be gleaned from the figure, at
sufficiently high $\ell$, shot noise dominates, especially in the
highest-redshift bins. We explore the dependence of the overall S/N on the
galaxy shot-noise contribution more below in Section~\ref{sec:galaxy_sn}.

Figure~\ref{fig:snr} shows the predicted S/N for the case of CMB-S4 and the
Roman HLS galaxies as a function of the galaxy window function parameters $z_0$
and $\Delta z$. In Figure~\ref{fig:kxg} we found that the cross-spectrum depends
relatively little on the extent of the galaxy window, while the galaxy angular
power spectrum in the cross-spectrum variance (i.e. in the denominator of
Equation~\ref{eqn:snr}) decreases as the width of the window increases. Hence,
the S/N for a single redshift bin is largest for a broad redshift window. In the
CMB-S4 case, we find that the highest S/N is attained for a redshift bin of
$\Delta z \sim 4.2$, centered on a redshift just a little bit lower than the
redshift of the signal peak.  In the case of our fiducial reionization history,
and CMB-S4 noise properties, the total S/N reaches 2.8$\sigma$ for a single,
broad ($\Delta z\sim 4.2$) redshift bin centered on $z_0\sim 9.2$. Given that
our forecasts are inevitably idealized, this seems somewhat marginal if
tantalizingly close to a detectable signal.

Note that given the wide parameter space of $z_0$ and $\Delta z$ considered,
there are some combinations that include contributions from the
post-reionization epoch ($z \leq 6$). Figure~\ref{fig:snr} shows a dashed line,
where combinations above this line would include such contributions. Rather than
include the post-reionization epoch and potentially double-count the low-$z$
portion of the signal, we truncate the window to have a minimum value of $z=6$,
both for the signal and noise components of the S/N. As such, the S/N in this
portion of the figure may not adequately capture the interplay between the
reionization-era kSZ and low-redshift kSZ signals. Nevertheless, this portion of
parameter space may still be useful with real data, as looking at
post-reionization redshift values can confirm that reionization is complete by
a particular redshift.

\begin{figure}[t]
  \centering
  \includegraphics[width=0.45\textwidth]{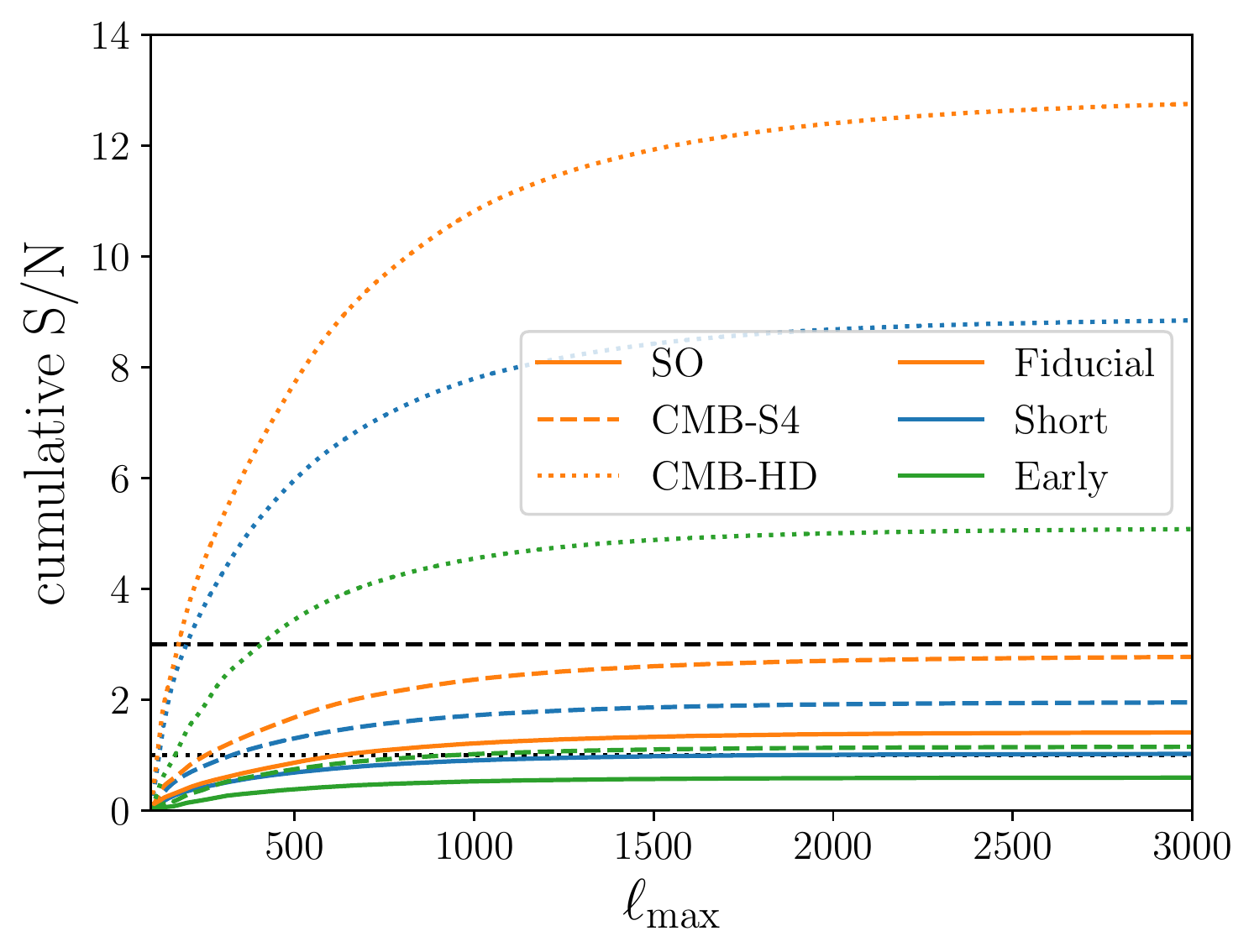}
  \caption{The cumulative S/N from Equation~(\ref{eqn:snr}) starting with a
    minimum of $\ell = 100$ and summed up to $\ell_\mathrm{max}$. For each line
    shown, we have chosen the galaxy window function parameters $z_0$ and
    $\Delta z$ that maximize the cumulative S/N. The dotted and dashed lines,
    respectively, indicate 1$\sigma$ and 3$\sigma$ detection levels. Most of the
    sensitivity comes from low-$\ell$ values ($\ell \lesssim 1000$).}
  \label{fig:snr_by_ell}
\end{figure}

Although this model is only borderline detectable with CMB-S4, the prospects are
more promising for the case of Roman HLS and CMB-HD
cross-correlations. Specifically, Figure~\ref{fig:snr_by_ell} shows the S/N
(summed over all $\ell$ modes) for different CMB surveys (each in combination
with the HLS galaxies) and our three reionization history models. In each case,
we consider a single redshift bin and choose the galaxy window parameters $z_0$
and $\Delta z$ that maximize the S/N. The first key conclusion here is that the
cross-spectrum signal is detectable at high (13$\sigma$) significance in our
fiducial model for CMB-HD. The signal in the short reionization history is
somewhat less detectable (8$\sigma$) because of the more rapid rise and fall of
the cross-spectrum in this scenario. The earlier reionization history is still
less detectable (4$\sigma$): Even though the signal is stronger in this model
(see Figure~\ref{fig:xhist_comp}), the peak occurs at higher redshifts where the
shot noise is larger. Note that the fiducial case is likely the most plausible
of these models given current observational constraints on the reionization
history. Hence, although these forecasts depend somewhat on the precise
reionization model adopted, the prospects of detecting this signal look quite
good for the case of CMB-HD.  See Table~\ref{table:snr} for the numerical values
in each reionization history and experiment.

Next, Figure~\ref{fig:snr_by_ell} further illustrates that most of the S/N comes
from modes with $\ell \lesssim 1000$, with the cumulative S/N rising steeply
between $500 \lesssim \ell \lesssim 1000$, especially in the most detectable
CMB-HD case. This cements our previous remarks that this statistic primarily
measures squeezed-triangle configurations. Finally, one can see that strong
detections appear a little out of reach for CMB-S4 ($\sim$3$\sigma$) and SO
($\sim$1$\sigma$) in our fiducial reionization history.

\subsection{From Detection to Characterization}
\label{sec:characterization}

In the preceding discussion, we focused primarily on the detection of the
overall signal, where we combined the total sensitivity across relatively wide
ranges in redshift and summed over all $\ell$ modes. However, the projected
sensitivity of CMB-HD is such that it becomes possible for us to move beyond a
simple detection and potentially characterize the evolution of the signal as a
function of redshift. Such an analysis would be useful for inferring the
reionization history, as Figure~\ref{fig:xhist_comp} shows that the signal
follows generic trends---at least across the three cases considered---with the
volume-averaged ionization fraction of the universe.

\begin{figure}[t]
  \centering
  \includegraphics[width=0.45\textwidth]{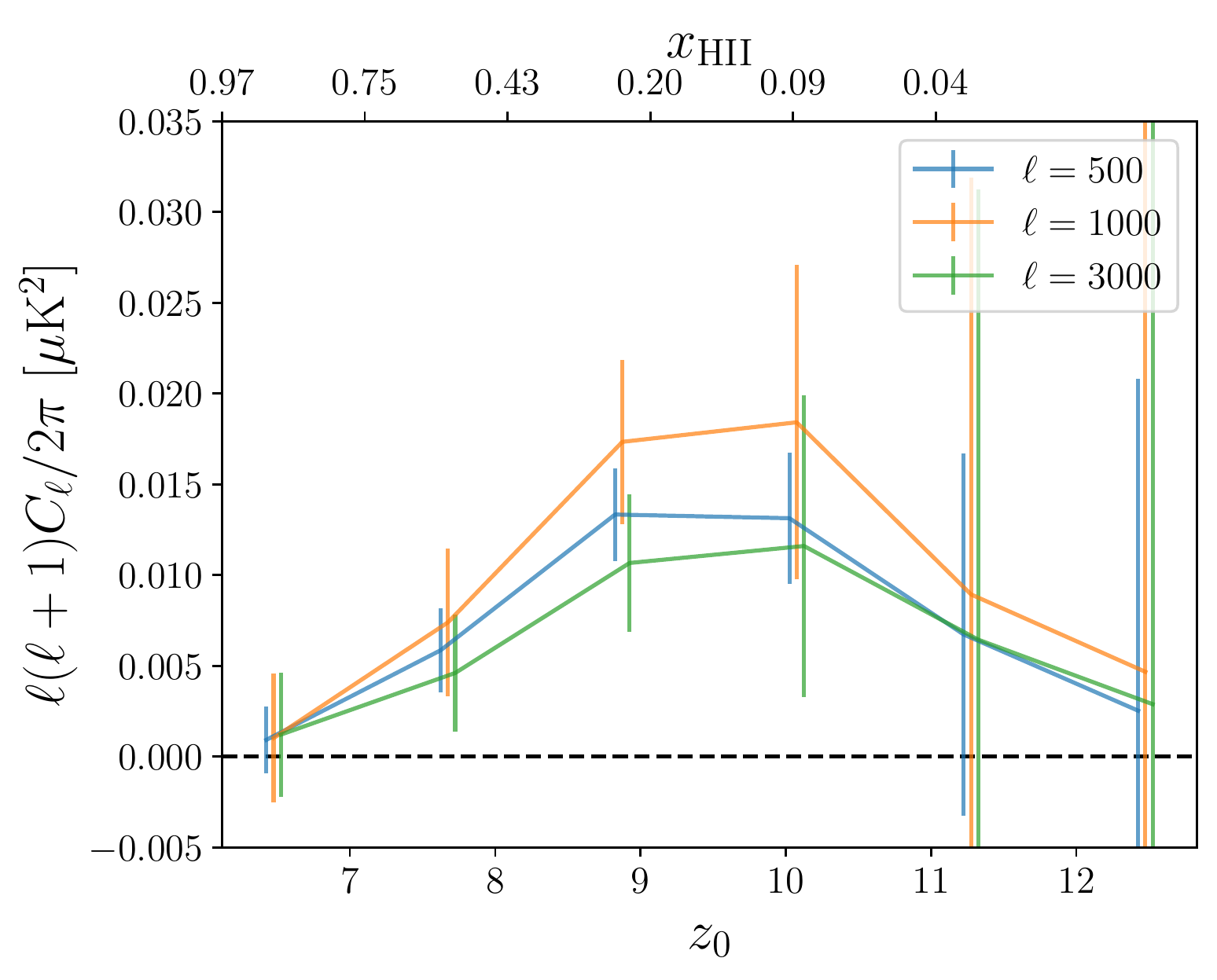}
  \caption{Forecast observational errors for the case of CMB-HD. This is similar
    to Figure~\ref{fig:kxg_of_z}, except here we show the expected error bars
    for a real planned survey rather than uncertainties on the simulation
    measurements. In the case of a future measurement using the planned CMB-HD
    survey and Lyman-break-selected galaxies from the Roman HLS, the
    characteristic redshift evolution of the signal may be measured with
    moderate statistical significance (see further discussion in
    Section~\ref{sec:characterization}).}
  \label{fig:real_noise}
\end{figure}

This is illustrated explicitly in Figure~\ref{fig:real_noise} which shows the
projected CMB-HD $\times$ HLS error bars across several independent photometric
redshift bins. Here we have binned the galaxy distribution into bins of width
$\Delta z =1$ and calculate the S/N (Equation~\ref{eqn:snr}) in a few different
$\ell$ ranges at each redshift. Specifically, we determine the S/N in three
bins, $250 \leq \ell \leq 750$, $750 \leq \ell \leq 1250$, and
$1500 \leq \ell \leq 4500$ centered on $\ell = \{500, 1000, 3000\}$,
respectively. As discussed previously, the expected photometric redshift
accuracy is $\sigma_z \sim 0.2$--0.3 and so we should be able to safely measure
the evolution across several independent $\Delta z=1$ redshift bins. Although
the error bars are large in the two highest-redshift bins, the remaining results
look promising for CMB-HD: One should be able to at least roughly detect the
peak in the signal and the subsequent decline in the kSZ$^2$--galaxy cross-power
spectrum as the universe is progressively reionized. Note that the different
nonoverlapping redshift bins here should be only weakly correlated with each
other.\footnote{In the Gaussian and Limber approximations, the covariance
  between CMB$^2$-galaxy cross-power spectra estimates in two redshift bins, $i$
  and $j$, with galaxy fluctuation fields $g_i$ and $g_j$, is
  $\mathrm{cov}\left[C^{T^2,g_i}_\ell; C^{T^2,g_j}_\ell \right]=
  \frac{C^{T^2,g_i}_\ell C^{T^2,g_j}_\ell}{(2 \ell + 1) f_{\rm sky}}$.  This
  formula assumes that the redshift bins are nonoverlapping. The covariance here
  is weak relative to the variance in the cross-spectrum estimates in each
  redshift bin.} From the error forecasts in Figure~\ref{fig:real_noise}, it is
thus clear that CMB-HD can measure the overall evolution of this signal as a
function of redshift, yielding important information regarding the timing and
duration of reionization.

\subsection{CMB Lensing Leakage}
\label{sec:lensing_leakage}

\begin{figure}[t]
  \centering
  \includegraphics[width=0.45\textwidth]{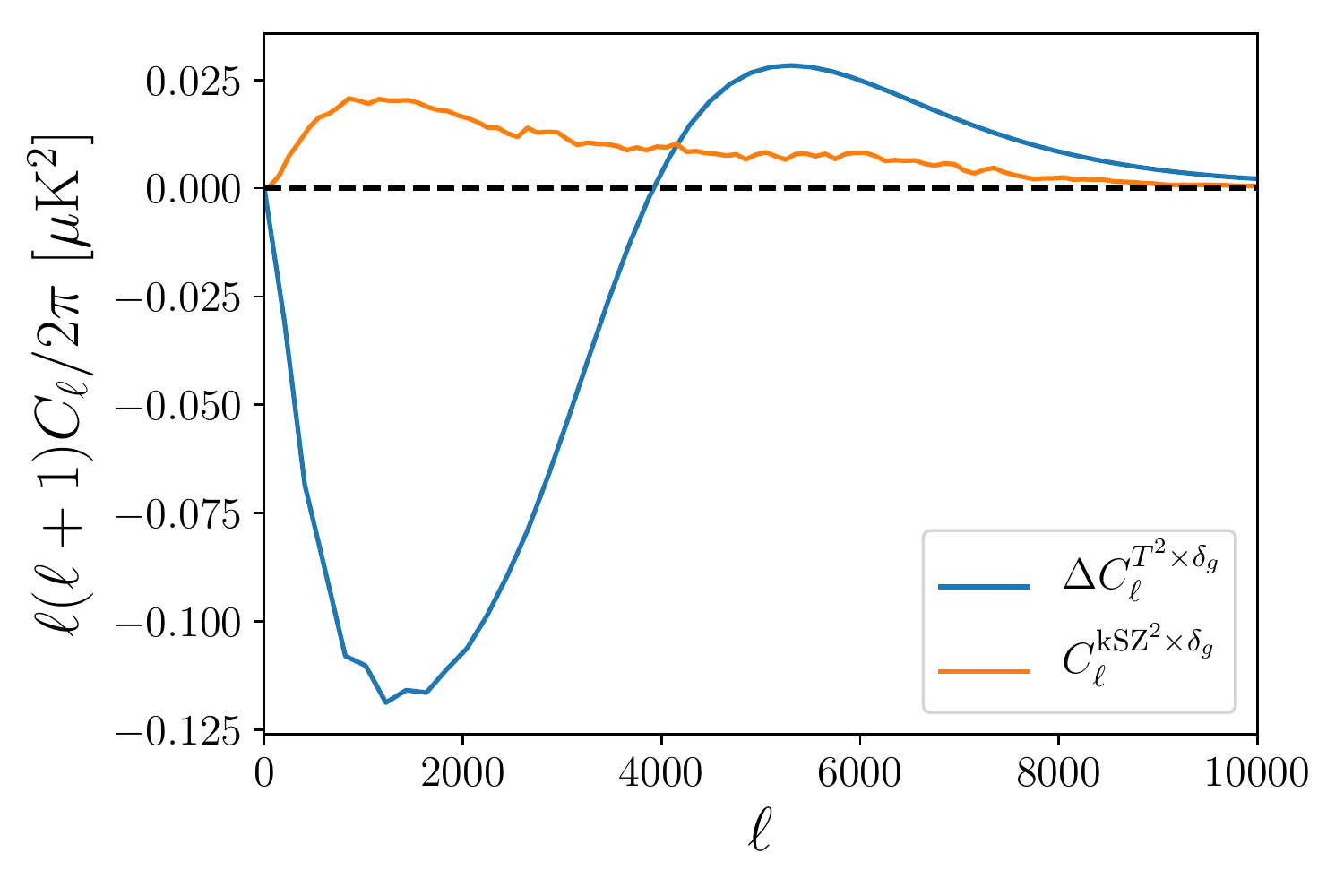}
  \caption{A comparison between the cross-correlation signal of interest
    $C_\ell^{\mathrm{kSZ}^2\times\delta_g}$ and the leakage due to CMB lensing
    $\Delta C_\ell^{T^2\times\delta_g}$. These results adopt the S4 filter and a
    galaxy window with $z_0=9.5$ and $\Delta z=1$. Although the lensing leakage
    is stronger in amplitude than the kSZ$^2$--galaxy cross-power spectrum
    signal, the $\ell$ dependence is very different, as in
    \citetalias{ferraro_etal2016}. It should therefore be possible to fit for
    both contributions without significantly increasing the errors on the
    kSZ$^2$--galaxy cross-power spectrum.}
  \label{fig:lensing}
\end{figure}

As discussed in previous works, e.g., \citetalias{ferraro_etal2016}, the
kSZ$^2$--galaxy cross-power spectrum estimates discussed here will be
contaminated by CMB lensing. The lensing effects must be separated from the
desired kSZ signal using its different angular dependence, which should be
distinctive and clean to model. Here we calculate the lensing contamination for
the very high-redshift application considered in the present work.

In order to investigate this, we determine the lensing leakage terms as
described in Section~B of \citet{kusiak_etal2021} (see their Equations
(9)-(15)). We ignore the magnification bias terms considered in that work
(their Section~C), which require modeling the full luminosity function of the
galaxy samples involved, although these would be interesting to consider in
future studies. In calculating the leakage, we adopt a linear-biasing model and
the linear theory matter power spectrum, which should be a good approximation for
our high-redshift application here.  As in previous work, the lensing leakage is
denoted by $\Delta C_\ell^{T^2\times\delta_g}$.

The resulting leakage is shown in Figure~\ref{fig:lensing}, as compared to the
cross-correlation signal of interest,
$C_\ell^{\mathrm{kSZ}^2\times\delta_g}$. This is for the case of a $\Delta z=1$
galaxy redshift bin, centered at $z_0=9.5$, close to the peak redshift for the
kSZ$^2$--galaxy cross-power spectrum. The lensing leakage shows a distinctive
anticorrelation at low $\ell$ and a positive correlation at higher $\ell$, as
discussed in \citetalias{ferraro_etal2016}.  Although we are considering very
high-redshift galaxy samples here, the CMB lensing leakage is still quite
important. This occurs in part because the CMB lensing kernel is quite broad,
and because although the matter distribution is relatively smooth at high
redshift, the Roman HLS galaxies are highly biased tracers of the matter
distribution. Indeed, the overall amplitude of the leakage is strongest around
$\ell \sim 1000$--2000 where it reaches
$\ell (\ell + 1) \Delta C_\ell^{T^2\times\delta_g}/(2 \pi) \sim -0.12 \mu {\rm
  K}^2$. This is a factor of $\sim 5$ larger in absolute value than the signal
we seek to measure.

Nevertheless, we believe this is not a big obstacle for our measurements. First,
the leakage has a very different $\ell$ dependence than the signal we seek to
measure. Therefore, one can hope to simultaneously fit for both components
without significantly inflating the error bar on the kSZ$^2$--galaxy
cross-spectrum. Second, the lensing contribution should be robust to model, up
to an overall bias factor which can be constrained independently from the
auto-power spectrum of the galaxy distribution itself. Third, the
kSZ$^2$--galaxy cross-power spectrum evolves strongly with redshift and the
patchy reionization contribution vanishes entirely once reionization completes,
below $z \lesssim 5.5$--6 or so. The lensing contribution evolves much less
sharply in redshift, especially in its $\ell$ dependence. That is, the differing
redshift evolution of the two contributions offers a further handle for
distinguishing them.  We will not quantify the prospects for joint fits further
here: While it is important to account for the lensing leakage, it should not be
a major impediment for measuring the kSZ$^2$--galaxy cross-power spectrum.  An
alternate approach is to reconstruct the projected mass distribution using
lensing-induced correlations between $E$- and $B$-mode polarization
signals. This polarization-only estimate can be used to determine the lensing
contribution to our kSZ$^2$--galaxy statistic without contamination from the kSZ
signal, which is intrinsically unpolarized \citepalias{ferraro_etal2016}.

\section{Further Improvements}
\label{sec:future_improvements}

Although the CMB-HD results above appear promising, it is useful to consider
possible directions for improving these measurements further. There are
essentially three areas where there is room for improvement. First, the S/N
forecasts are sensitive to residual foregrounds in the CMB maps, and hence
further foreground mitigation may help. Second, a stronger detection is possible
if a larger common region of the sky can be covered by both the galaxy and CMB
surveys. Finally, a deeper galaxy survey would reduce the shot noise and boost
the overall detection significance.

\subsection{CMB Noise Models}
\label{sec:cmb_noise}

In the calculations thus far, we have included estimates of the residual
foreground contributions to the effective noise variance after harmonic-based
ILC component separation. In general, the resulting variance is substantially
larger than the purely instrument-based noise power spectrum (described by
Equation~\ref{eqn:beam}). In order to quantify the impact of foregrounds, we
consider here the idealized pure instrumental-noise-dominated limit in which
residual foregrounds are absent.

\begin{deluxetable*}{@{\extracolsep{6pt}}cccccc}
\tablecaption{Cumulative S/N For Experiments
\tablewidth{0pc}
\label{table:snr}}
\tablehead{\colhead{Experiment} & \colhead{History} & \colhead{ILC Noise} & \colhead{Instrument} & \colhead{ILC Noise} & \colhead{Instrument} \\[-1em]
\colhead{} & \colhead{} & \colhead{} & \colhead{Noise} & \colhead{} & \colhead{Noise} \\
\colhead{} & \colhead{} & \multicolumn{2}{c}{With shot noise} & \multicolumn{2}{c}{Without shot noise}
}
\startdata
SO     & Fiducial & 1.4  & 1.7  & 2.4 & 2.9 \\
       & Short    & 1.0  & 1.3  & 1.7 & 2.0 \\
       & Early    & 0.60 & 0.77 & 2.0 & 2.4 \\ %\hline
       \cline{1-2} \cline{3-4}\cline{5-6}
CMB-S4 & Fiducial & 2.8  & 11  & 4.8 & 18 \\
       & Short    & 2.0  & 8.3 & 3.1 & 12 \\
       & Early    & 1.2  & 4.8 & 3.8 & 15 \\ %\hline
       \cline{1-2} \cline{3-4}\cline{5-6}
CMB-HD & Fiducial & 13  & 35  & 23 & 64 \\
       & Short    & 8.9  & 25 & 14 & 40 \\
       & Early    & 5.1  & 13 & 17 & 46 \\
\enddata
\tablecomments{The cumulative S/N for different experiments. We show the results
  for each ionization history, choosing the values for the galaxy selection
  function $W_g$ that maximize the S/N, for the cases of using the post-ILC
  noise level or just the nominal instrumental noise, summed over all $\ell$
  modes. Further discussion is in Section~\ref{sec:cmb_noise}.  We also include
  or ignore the effects of shot noise in the galaxy survey (see
  Section~\ref{sec:galaxy_sn}.)}
\end{deluxetable*}

The first and second columns of Table~\ref{table:snr} show the cumulative S/N we
forecast for the different experiments and noise models. We include both the
post-ILC noise levels for each experiment, as well as the naive instrument-only
noise. For each experiment, including only the instrumental noise (i.e.,
assuming the foreground contamination can be removed entirely) increases the
S/N. Although the difference is relatively modest for SO, the gain is much more
substantial for CMB-S4 and CMB-HD. The S/N increases roughly by a factor of 4
for CMB-S4 and by a factor of around 3 for CMB-HD. The larger boosts from
CMB-S4 and CMB-HD arise because of the improved sensitivity and angular
resolution of these experiments relative to SO. This larger improvement factor
of about 4 for CMB-S4 compared to about 3 for CMB-HD owes to a greater
difference in the power spectrum of the filtered CMB-squared map
$C_\ell^{\bar{T}^2\bar{T}^2,f}$ at low-$\ell$ modes, where most of the S/N
arises for each instrument. In the case of CMB-HD, the cumulative S/N forecasted
in the noise-dominated limit is 33: This is an impressive detection, which would
allow a fairly detailed characterization of the cross-spectrum signal along the
lines of, yet better than, Figure~\ref{fig:real_noise}.

Although it will be hard to remove the foregrounds down to below the
instrumental-noise limit, note that the high sensitivity and angular resolution
of CMB-HD and CMB-S4 may allow additional mitigation steps beyond the ILC
component separation considered here. In particular, bright infrared sources and
radio galaxies may be identified directly in the CMB maps and masked. This would
help to reduce the level of residual CIB and radio galaxy foregrounds in the
maps, and push toward the noise-dominated sensitivity limit, although a
detailed investigation is beyond the scope of this paper. In principle, the CMB
surveys might allocate additional observing time to the HLS field to help with
bright foreground source excision on the cross-correlation field. Another
futuristic possibility is to employ additional frequency channels to aid
component-separation efforts. Along these lines, we have also considered the
prospects of measuring the cross-power spectrum with the proposed space-based
Probe of Inflation and Cosmic Origins experiment (PICO;
\citealt{pico2019}). Although it has 21 frequency channels, which help with the
foreground separation, it has coarser angular resolution and we find that it
will be unable to detect our cross-spectrum statistic.

\subsection{Galaxy Survey Sky Coverage and Depth}
\label{sec:galaxy_sn}

The other potential axes for improving the S/N involve the sky coverage and
survey depth of the galaxy survey. First, let us consider the sky coverage. The
relevant sky area here is, of course, the common region of sky surveyed by both
the galaxy survey and the CMB experiments. In practice, the CMB surveys plan to
cover very large fractions of the sky (with coverage of
$f_\mathrm{sky} \sim 0.4$ or larger), and so the main limitation here is the
coverage of the HLS. At fixed galaxy density, the S/N scales simply with the
square root of $f_{\rm sky}$ (Equation~(\ref{eqn:snr})), and so it is easy to
quantify the potential benefit here. Specifically, the sky coverage of the HLS
is $f_{\rm sky} = 0.05$, and so in the limiting (yet hard to achieve in
practice) case that one could cover the entire sky at the HLS depth, the S/N
boost would be a factor of $\sqrt{20}$. Combined with CMB-HD--type sensitivity,
the total S/N in this case would reach 50 for post-ILC noise or 150 in the
instrumental-noise-dominated limit.

In order to quantify the impact of shot noise, we can again take the idealized
limit that only sample variance in the galaxy distribution contributes to the
cross-spectrum variance. In order to make this estimate, we assume that the
clustering bias remains constant as the shot noise is reduced to zero. Note that
at least in the linear-biasing- and sample-variance-dominated regime,
Equation~(\ref{eqn:snr}) is independent of $b_g$ and so our conclusions should
be insensitive to this simplifying assumption. In this case, we find that the
total S/N forecast with CMB-HD is 22 for the post-ILC case and 62 for the
instrumental-noise limit. In the ultimate, full-sky, negligible foreground and
shot-noise limit with CMB-HD level instrumental noise, the S/N would reach 280.

Although these simplified limiting cases are illustrative, a more nuanced
question relates to whether one can improve the S/N at fixed total observing
time for the galaxy survey. In other words, for the purposes of optimizing the
S/N of the cross-spectrum considered here, would it be better to adopt a deeper
or wider observing strategy for the HLS? A detailed exploration of this is
beyond the scope of the present work, but here we simply note that at
$\ell \lesssim 1000$ the galaxy abundance fluctuations are mainly dominated by
clustering rather than shot noise (Figure~\ref{fig:galaxy_cl}). This suggests
that a more optimal strategy would be to cover a wider region of the sky, less
deeply, until the clustering and shot noise become comparable for the most
important $\ell$ modes. This is unlikely, however, to boost the expected S/N
substantially.

\section{Conclusion}
\label{sec:conclusion}

We have modeled the kSZ$^2$--galaxy cross-power spectrum signal during the
EoR. The usual CMB angular power spectrum approaches for detecting the patchy
reionization kSZ signal require templates for both the patchy reionization and
late-time kSZ signals. Here, the post-reionization kSZ effect only contributes
to the cross-correlation noise and does not produce an average bias and so this
method may allow one to directly extract patchy reionization contributions to
the CMB anisotropies.  Furthermore, we find that the signal -- as measured in
different photometric galaxy redshift bins -- evolves in a distinctive way
across the EoR. Specifically, the signal peaks at an amplitude of
$C_\ell^{\mathrm{kSZ}^2\times\delta_g} \sim 0.02$~$\mu$K$^2$ (after suitable
filtering) for $\ell$ modes around $500 \lesssim \ell \lesssim 3000$ when the
volume-averaged ionization fraction is around $x_{\rm HII} = 0.15$--0.20.

These predictions rely somewhat on the seminumeric \texttt{zreion} model
introduced in \citet{battaglia_etal2013a}. Although these models produce
ionization fields that are in reasonable agreement with more-detailed radiation
hydrodynamical simulations of reionization, they should be further cross-checked
for the statistics considered here.

We forecast that a 12$\sigma$ detection of this signal is possible using future
cross-correlation measurements with CMB-HD and Lyman-break galaxies from the
Roman HLS. Furthermore, we expect to be able to detect the distinctive redshift
evolution in our models at moderate statistical significance. In the case of
CMB-S4, however, we find only a 2.7$\sigma$ detection significance and a
1.4$\sigma$ significance for SO, each in correlation with the Roman HLS
sample. Although the prospects are marginal for these upcoming experiments, they
may provide bounds on the signal strength and test for unanticipated
possibilities, while providing a real-world test of the technique discussed
here.

Furthermore, it may be possible to extend and generalize the approach discussed
here. Specifically, we model the kSZ$^2$--galaxy cross-power spectrum in this
work mainly because it has been successfully applied to real data in the
post-reionization universe \citep{kusiak_etal2021} and by virtue of its
simplicity. However, it may be interesting to generalize this and calculate the
full kSZ--kSZ--galaxy bispectrum for a broad range of triangle
configurations. This may boost the detection sensitivities and allow
measurements in advance of CMB-HD.

\begin{acknowledgments}
  We thank J. Colin Hill and Simone Ferraro for helpful discussions regarding
  ILC noise, and we thank James Aguirre for insight on CMB observing
  strategies. We also thank the anonymous reviewer for valuable feedback on this
  manuscript.  P.L. acknowledges support from the Berkeley Center for Cosmological
  Physics. J.S. and A.L. acknowledge support from NASA ATP grant 80NSSC20K0497. This
  work used the Extreme Science and Engineering Discovery Environment (XSEDE),
  which is supported by National Science Foundation grant number ACI-1548562
  \citep{xsede2014}. Specifically, it used the Bridges-2 system, which is
  supported by NSF award number ACI-1445606, at the Pittsburgh Supercomputing
  Center (PSC, \citealt{bridges2015}).
\end{acknowledgments}

\bibliography{mybib}
\bibliographystyle{aasjournal}

\end{document}